\documentclass[sigconf]{acmart}

\hypersetup{
    pdfauthor={Chao Zhang},
    pdfsubject={CHI 2024},
    pdftitle={Mathemyths: Leveraging Large Language Models to Teach Mathematical Language through Child-AI Co-Creative Storytelling},
    pdfkeywords={CHI 2024}
}

\usepackage{subcaption}
\usepackage{tabularx}

\usepackage{xcolor}
\usepackage{soul}
\soulregister{\cite}7
\soulregister{\citep}7
\soulregister{\citet}7
\soulregister{\ref}7
\soulregister{\pageref}7

\usepackage{wrapfig}
\usepackage{siunitx,etoolbox}
\usepackage{dcolumn,booktabs}
\usepackage{multirow}
\usepackage{float}
\newcolumntype{d}[1]{D{.}{.}{#1}}
\usepackage{bm}
\newrobustcmd*{\bftabnum}{%
	\bfseries
	\sisetup{output-decimal-marker={\textmd{.}}}%
}

\usepackage{array}
\newcolumntype{C}[1]{>{\centering\arraybackslash}p{#1}}

\usepackage{longtable}

\AtBeginDocument{%
  \providecommand\BibTeX{{%
    \normalfont B\kern-0.5em{\scshape i\kern-0.25em b}\kern-0.8em\TeX}}}

\copyrightyear{2024}
\acmYear{2024}
\setcopyright{acmlicensed}\acmConference[CHI '24]{Proceedings of the CHI Conference on Human Factors in Computing Systems}{May 11--16, 2024}{Honolulu, HI, USA}
\acmBooktitle{Proceedings of the CHI Conference on Human Factors in Computing Systems (CHI '24), May 11--16, 2024, Honolulu, HI, USA}
\acmDOI{10.1145/3613904.3642647}
\acmISBN{979-8-4007-0330-0/24/05}

\newcommand{\eg}{{e.g.,\ }}

\newcommand{\ie}{{i.e.,\ }}

\newcommand{\TOOL}{{\textsc{Mathemyths}\ }}

\definecolor{oxfordblue}{rgb}{0.0, 0.13, 0.28}
\definecolor{harvardcrimson}{rgb}{0.79, 0.0, 0.09}
\definecolor{dartmouthgreen}{rgb}{0.05, 0.5, 0.06}
\definecolor{princetonorange}{rgb}{1.0, 0.56, 0.0}
\definecolor{yaleblue}{rgb}{0.06, 0.3, 0.57}
\definecolor{usccardinal}{rgb}{0.6, 0.0, 0.0}
\definecolor{uclablue}{rgb}{0.33, 0.41, 0.58}
\definecolor{msugreen}{rgb}{0.09, 0.27, 0.23}
\definecolor{cornellred}{rgb}{0.7, 0.11, 0.11}
\definecolor{pomegranate}{RGB}{192, 57, 43}
\definecolor{anti-pomegranate}{RGB}{43,178,192}
\definecolor{alizarin}{RGB}{231, 76, 60}
\definecolor{anti-belize}{RGB}{185, 41, 56}
\definecolor{belize}{RGB}{41, 128, 185}
\definecolor{peter}{RGB}{52, 152, 219}
\definecolor{green}{RGB}{22, 160, 133}
\definecolor{anti-green}{RGB}{160,22,118}
\definecolor{turquoise}{RGB}{26, 188, 156}
\definecolor{pumpkin}{RGB}{211, 84, 0}
\definecolor{anti-pumpkin}{RGB}{0,22,211}
\definecolor{carrot}{RGB}{230, 126, 34}
\definecolor{wisteria}{RGB}{142, 68, 173}
\definecolor{anti-wisteria}{RGB}{99,173,68}
\definecolor{amethyst}{RGB}{155, 89, 182}
\definecolor{nephritis}{RGB}{39, 174, 96}
\definecolor{anti-nephritis}{RGB}{174,39,117}
\definecolor{grey-bg}{RGB}{242,242,235}

\newcommand{\revision}[1]{{#1}}
\newcommand{\correct}[1]{{#1}}

\begin{document}

\title[Mathemyths]{Mathemyths: \revision{Leveraging Large Language Models to} Teach Mathematical Language through Child-AI Co-Creative Storytelling}

%

\author{Chao Zhang}
\authornote{This work was carried out when the author was a research intern at the University of Michigan.}
\email{cz468@cornell.edu}
\orcid{0000-0003-4286-8468}
\affiliation{%
 \institution{Cornell University}
 \city{Ithaca, NY}
 \country{USA}}

\author{Xuechen Liu}
\email{liuxuech@umich.edu}
\orcid{0009-0003-7847-5604}
\affiliation{%
 \institution{University of Michigan}
 \city{Ann Arbor, Michigan}
 \country{USA}}

\author{Katherine Ziska}
\email{kziska@umich.edu}
\orcid{0009-0002-9198-9936}
\affiliation{%
 \institution{University of Michigan}
 \city{Ann Arbor, Michigan}
 \country{USA}}

\author{Soobin Jeon}
\email{sbjeon@umich.edu}
\orcid{0009-0009-5470-4576}
\affiliation{%
 \institution{University of Michigan}
 \city{Ann Arbor, Michigan}
 \country{USA}}

\author{Chi-Lin Yu}
\email{chilinyu@umich.edu}
\orcid{0000-0002-4381-7163}
\affiliation{%
 \institution{University of Michigan}
 \city{Ann Arbor, Michigan}
 \country{USA}}

\author{Ying Xu}
\email{yxying@umich.edu}
\orcid{0000-0003-2079-2394}
\affiliation{%
 \institution{University of Michigan}
 \city{Ann Arbor, Michigan}
 \country{USA}}
\renewcommand{\shortauthors}{Chao Zhang et al.}

%
\begin{abstract}

Mathematical language is a cornerstone of a child's mathematical development, and children can effectively acquire this language through storytelling with a knowledgeable and engaging partner. In this study, we leverage the recent advances in large language models to conduct free-form, creative conversations with children. Consequently, we developed \textsc{Mathemyths}, a joint storytelling agent that takes turns co-creating stories with children while integrating mathematical terms into the evolving narrative. This paper details our development process, illustrating how prompt-engineering can optimize LLMs for educational contexts. Through a user study involving 35 children aged 4-8 years, our results suggest that when children interacted with \textsc{Mathemyths}, their learning of mathematical language was comparable to those who co-created stories with a human partner. However, we observed differences in how children engaged with co-creation partners of different natures. Overall, we believe that LLM applications, like \textsc{Mathemyths}, offer children a unique conversational experience pertaining to focused learning objectives.
\end{abstract}



\begin{teaserfigure}
 \includegraphics[width=\linewidth]{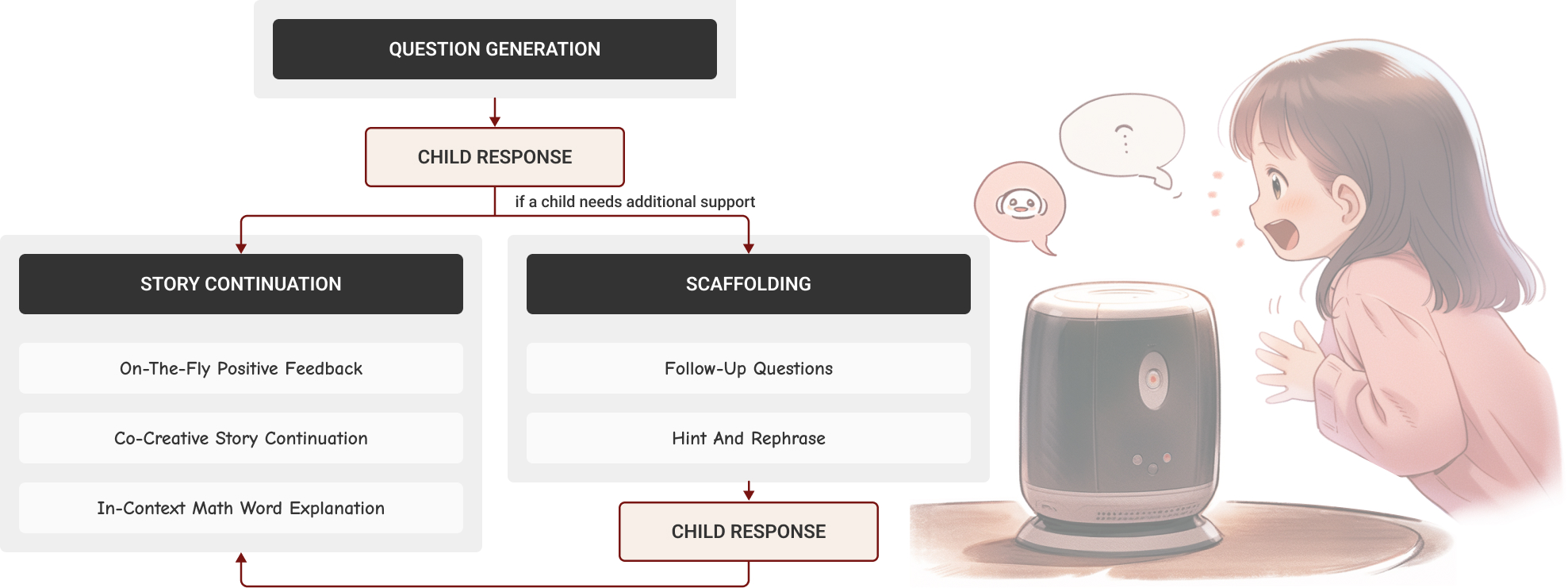}
 \caption{\revision{Interaction between a child and \textsc{Mathemyths}: demonstrating the system's ability to teach mathematical language through child-AI co-creative storytelling. \textsc{Mathemyths} provides open-ended questions to solicit how the child wishes the story should progress, on-the-fly feedback to acknowledge the child's responses, and co-creative story continuation with in-context explanations of math words. When the child needs additional support to continue the story, \textsc{Mathemyths} offers scaffolding through follow-up questions and ``hint \& rephrase'' strategies.}}
 \Description{Interaction between a child and Mathemyths: demonstrating the system's ability to teach mathematical language through child-AI co-creative storytelling. Mathemyths provides open-ended questions to solicit how the child wishes the story should progress, on-the-fly feedback to acknowledge the child's responses, and co-creative story continuation with in-context explanations of math words. When the child needs additional support to continue the story, Mathemyths offers scaffolding through follow-up questions and ``hint and rephrase'' strategies.}
 \label{teaser_figure}
\end{teaserfigure}

\maketitle

\section{Introduction}

Mathematics is a fundamental building block in children's early education and is a strong predictor of their future achievements~\cite{bynnerDoesNumeracyMatter1997,duncanSchoolReadinessLater2007,parsonsDoesNumeracyMatter2005a,saramaEarlyChildhoodMathematics2009,starkeyEnhancingYoungChildren2004}. Particularly in early childhood, mathematical learning is not just about learning to count or calculate; it also emphasizes the ability to understand and apply mathematical language, such as ``half'' or ``equal''. 
\revision{Children often learn these words through everyday conversations with others~\cite{thippana2020parents}. For instance, when a parent asks, ``Do you have more cookies than me?'' children are introducing the quantitative term ``more.'' However, studies have found that the amount and quality of math language children \correct{encounter} through these everyday, spontaneous conversations can vary greatly, which often aligns with socioeconomic differences ~\cite{dearing2022socioeconomic}. This variation may have translated into disparities in early math abilities ~\cite{elliott2018ses}. }\looseness=-1

\revision{To enrich \correct{children's} experience of math language beyond everyday conversations, researchers have emphasized the importance of including a language component in general mathematics instruction, which allows children to engage with math language \correct{more systematically}. However, it is challenging to teach math language abstractly to young children ~\cite{litkowski2020preschoolers}. One of \correct{the approaches} widely used with \correct{preschoolers} and early elementary school children is to embed math language within the context of \textit{narrative stories}, making abstract language more tangible and engaging for young learners \cite{purpuraCausalConnectionsMathematical2017,hassinger-dasReadingStoriesLearn2015}.} Typically, this story-based learning approach involves encouraging children to engage in dialogues within a storyline, where they understand and practice math terms with guidance from a more knowledgeable individual, such as a parent, teacher, or older peer ~\cite{purpuraCausalConnectionsMathematical2017,hassinger-dasReadingStoriesLearn2015,busJointBookReading1995,durantiRethinkingContextLanguage1992a,morrowRetellingStoriesStrategy1985}. Yet, the practicality of teaching math language through joint storytelling is debatable, as it demands a one-on-one engaged storytelling session that is logistically infeasible given the limited resources most educational institutions face ~\cite{purpuraDevelopmentMathematicalLanguage2019,purpuraCausalConnectionsMathematical2017,purpuraEngagingCaregiversChildren2021}.\looseness=-1

In response to this challenge, researchers have long been investigating AI's potential to simulate human-like storytelling interactions~\cite{chungTaleBrushSketchingStories2022,osoneBunChoAISupported2021a,leeInteractiveChildrenStory2022,osoneBunChoAISupported2021a,yuanWordcraftStoryWriting2022,shakeriSAGACollaborativeStorytelling2021}. 
However, before the emergence of Large Language Models (LLMs), AI was primarily optimized for structured interactions with children, such as asking questions and providing evaluative feedback~\cite{ruanSupportingChildrenMath2020, xuElinorTalkingMe2022}. 
It is with the recent advancements in LLMs, notably GPT-4~\cite{GPT4}, that the technical difficulty of facilitating spontaneous and creative conversations has started to see promising developments~\cite{kasneciChatGPTGoodOpportunities2023}. 
However, the feasibility and efficacy of utilizing LLMs as partners to engage with children in storytelling remains unclear, especially in the context of supporting the learning of math language. Moreover, from a design perspective, it is unclear what specific workflow is necessary to tailor LLMs' capacities to a younger audience ~\cite{weiEmergentAbilitiesLarge2022}.

This study outlines the development process for creating \textsc{Mathemyths}
, a novel co-creative storytelling system powered by a GPT-based conversational agent, with the goal of teaching math language to children (Fig.~\ref{teaser_figure}). We then present empirical findings from a user study comparing learning and engagement outcomes using \textsc{Mathemyths} to those of a comparable human-guided interaction. Specifically, we aim to answer the following three research questions:\looseness=-1 

\begin{enumerate}
  \item[\textit{\textbf{RQ1}}] \textit{How can prompt engineering be employed to optimize LLMs as story co-creators tailored for children's educational purposes?}
  \item[\textit{\textbf{RQ2}}] \textit{Do interactions with LLMs effectively engage children and provide an enjoyable experience during the co-creation of stories?}
  \item[\textit{\textbf{RQ3}}] \textit{Can children's such interactions with LLMs result in comparable learning outcomes seen in human-child storytelling interactions?}
\end{enumerate}

Our results indicate that when children interacted with \textsc{Mathemyths}, their learning of mathematical language appeared to be comparable to those who co-created stories with a human partner. 
\revision{However, while both younger and older children benefited from interacting with \textsc{Mathemyths}, we observed differences in the ways children from different age groups engage with co-creation partners of different natures.} Overall, we believe that LLM applications, such as \textsc{Mathemyths}, can offer children a unique conversational experience, complementing their interactions with others.
Our findings have implications for the design of LLM-based applications aimed at supporting children's learning. 
Additionally, they highlight areas requiring further consideration to ensure the technology is both supportive and effective for diverse learning experiences. 

\looseness=-1

\section{Related Work}
In this section, we review research on (1) teaching math language through storytelling, (2) conversational interfaces for children, and (3) using LLMs for child-facing conversational interfaces.

\subsection{Teaching Math Language through Storytelling}


Children's mathematical abilities are significantly influenced by their language skills~\cite{morrowRetellingStoriesStrategy1985, parsonsDoesNumeracyMatter2005a, sunCollaborativeStorytellingRobot2017, wongSpokenNarrativeAssessment2016, wuImprovingVocabularyAcquisition2008, xuDialogueConversationalAgent2022,  zhangStoryBuddyHumanAICollaborative2022}. 
Indeed, the Common Core State Standards, a US educational framework, highlight the need for students to be able to understand and apply specific mathematical language to describe the math concepts ~\cite{akkus2016common}. For instance, by second grade, students are expected to b e familiar with terminology associated with fundamental math operations: addition, subtraction, multiplication, and division. Nevertheless, while some of this math language can be replaced by informal, everyday expressions (e.g., ``putting together'' as opposed to ``addition'') to convey a similar meaning, equipping students with precise math terminologies is still necessary. Such terminology serves as a bridge to help children transition from their concrete everyday way of thinking to more abstract analytical thinking.

\revision{Children engage in math language in different ways, including their daily dialogues with teachers and parents~\cite{thippana2020parents}. Research has indicated a positive correlation between the extent and frequency of math language used by parents or teachers and the development of children's math abilities at the onset of elementary education~\cite{purpuraEngagingCaregiversChildren2021}. However, the extent of math talk to which children are exposed varies significantly, likely influenced by the differing levels of awareness, expertise, or comfort among parents and teachers ~\cite{elliott2017understanding, dearing2022socioeconomic}. This variation is recognized as a contributing factor to the disparities observed in children's math abilities ~\cite{elliott2018ses}. Furthermore, while earlier studies predominantly focused on math language produced by parents and teachers, with children as recipients, more recent research has broadened this perspective. It emphasizes the opportunities for children to not only be exposed to but also generate math talk themselves~\cite{ainaImportanceOralStorytelling1999,schiroOralStorytellingTeaching2004}. Interestingly, these studies have revealed that the children's own mathematical conversations play a crucial, if not more significant, role in shaping their mathematical competence, sometimes even outweighing the influence of the math talk provided by parents or teachers~\cite{xuMathKingdomTeachingChildren2023b,turanMathematicalLanguageMathematical2022}. }

\revision{To address the varying degrees of math talk children encounter in spontaneous circumstances and to encourage their active engagement in math discussions, researchers have developed various interventions.} 
These interventions often utilize a storytelling approach~\cite{glaserNarrativebasedLearningPossible2009,hassinger-dasReadingStoriesLearn2015a,hojnoskiEmbeddingMathematicalDialogue2014,jenningsIncreasingInterestAchievement1992a,powellInfluenceMathematicsVocabulary2015}, where children, along with their parents or teachers, engage in math-themed narratives. Discussion prompts are provided within these narratives to encourage children to use math language. 
For example, Purpura et al.~\cite{purpuraCausalConnectionsMathematical2017} \revision{devised an eight-week intervention in Head Start classrooms, targeting children aged three to five. Researchers read six storybooks containing math vocabulary (e.g., ``fewer'') to the children. During the readings, researchers asked children questions that expanded upon the book's existing math language (e.g., 'How do we know there were a lot and not just a few?'). Additionally, if a child expressed confusion about a math term, the researchers explicitly defined and explained it. The researchers found that children who participated in the intervention significantly outperformed those in a comparison group, who continued with business as usual, not only in a math language assessment but also in a math knowledge assessment. Another study used a similar story-based approach among kindergarten children with numeracy difficulties ~\cite{hassinger-dasReadingStoriesLearn2015a}. This study found that the students who engaged in the stories and related dialogue learned math vocabulary better than the other group of students who received direct instruction of the math vocabulary.

Traditionally, many story-based approaches have incorporated dialogue into existing narratives, thus, to some extent, constraining children's creative involvement in shaping the stories. Thus, the research community has also been promoting an approach that encourages children to contribute and decide how they want the narrative storyline to progress, which is believed to empower children to bring their lived experiences to the dialogue to support their meaning making ~\cite{nelson1989storytelling}. This type of child-led storytelling has been adopted in preschool and early elementary classrooms and has proven to be feasible and age-appropriate. For instance, Flynn's study focused on a small group of four- to five-year-old children who were co-creating stories alongside an adult facilitator ~\cite{flynn2018ideas}. The research suggested that these children were able to meaningfully advance the story plot, drawing inspiration from ideas contributed by their peers. It should be noted that, facilitation from an adult is important for children to fully engage in this process.} In particular, the study highlighted the facilitator's role in posing questions to guide children's responses and offer constructive feedback responsive to children's contributions. Such \textit{question-feedback-scaffolding} could potentially lower the cognitive demand required for certain challenges children face when continuing the story~\cite{xuElinorTalkingMe2022}. \revision{Building on this concept, the HCI community has long been exploring technology-mediated platforms to support children storytelling. We will discuss this line of prior research in detail in the section below. }

\subsection{Conversational Interfaces for Children}

With the rapid advancement of AI, conversational user interfaces (CUIs) can now simulate interpersonal interactions through natural spoken language with children~\cite{porcheronVoiceInterfacesEveryday2018}. This type of speech-based interaction may not require children to have reading and typing skills, as is the case with other graphical user interfaces, thus removing the barrier for children who have not yet fully developed their literacy skills. Furthermore, it eliminates the need for a screen, potentially reducing tensions between children and their parents regarding screen time\cite{hinikerScreenTimeTantrums2016,kawasWhenScreenTime2021a}.
Previous research has identified two prevalent types of interactions that children tend to have with CUIs~\cite{xuAreCurrentVoice2021}: voice assistants supporting open-domain conversations (e.g., Apple Siri, Amazon Alexa, Google Assistant) and voice-based apps that specifically designed for children for domain-specific conversations~\cite{leeDAPIEInteractiveStepbyStep2023,beneteauParentingAlexaExploring2020a,bentleyUnderstandingLongtermUse2018,lovatoHeyGoogleUnicorns2019,lovatoYoungChildrenVoice2019}. This type of specifically design apps have been found to support children's learning across a number of disciplines, including math ~\cite{hoDesigningParentchildrobotInteractions2023,xuMathKingdomTeachingChildren2023b}, science~\cite{xuElinorTalkingMe2022}, computational thinking~\cite{dietzStoryCoderTeachingComputational2021}, and literacy ~\cite{xuAreCurrentVoice2021}.
\revision{Though varied in their
specific design, these applications are grounded in a common principle by utilizing conversational AI as a language partner to engage children in discussions related to the relevant topics. } \looseness=-1

For instance, Ho and colleagues developed a social robot that provides math-related prompts in storybooks for parents of four- and five-year-olds, covering concepts like subtraction, addition, and comparisons ~\cite{hoDesigningParentchildrobotInteractions2023}. The study revealed that parents found the robot's questions inspiring and that these questions led to more meaningful conversations with their children. Another example is a voice-based game designed to teach children aged four to seven ~\cite{xuMathKingdomTeachingChildren2023b}. In this game, children interacted with visual representations of shapes in different colors and sizes, receiving prompts such as ``turn the tallest triangle into a circle.'' The game also featured back-and-forth questions to guide children through this process. A user study involving 18 children demonstrated that this game helped them learn the target math terms and engage in verbal reasoning about their decisions. \revision{Although these two studies utilized either the Wizard-of-Oz approach or pre-scripted dialogues, they provided insights and evidence of feasibility in terms of how CUIs could be designed to support math talk among preschool and early elementary-aged children. These insights have paved the way for our current research, which seeks to enable free-flowing, story-based interactions for children in the context of math dialogues.}\looseness=-1


\revision{Another related research area is focused CUI and storytelling specifically, where an agent collaboratively co-creates stories with children ~\cite{zhangStoryDrawerChildAI2022,zhangStoryDrawerCoCreativeAgent2021,xuDialogueConversationalAgent2022,xuExploringYoungChildren2020,wuImprovingVocabularyAcquisition2008}, mirroring the common childhood activity children have with teachers or parents.} Among these studies, a prevailing model involves the agent listening to a child's stories and then periodically offering generic, template-based response that are not usually responsive to children's specific contribution ~\cite{bersInteractiveStorytellingSystems1998,ryokaiVirtualPeersPartners2003,westlundInterplayRobotLanguage2015}. For instance, Bers and colleague developed a plush rabbit that reacts to a child's storytelling by moving its ears and shaking its body ~\cite{bers1998interactive}. When the child pauses, the rabbit then prompts them with scripted follow-up questions to encourage further storytelling contributions \cite{bersInteractiveStorytellingSystems1998}. This interactive strategy has been utilized in more recent studies as well ~\cite{hubbard2021child}. \looseness=-1

While these studies have found that even this limited and structured interactivity can support children's storytelling, the agent made in fact minimal contributions to children's stories. Xu and colleagues referred to this type of interaction as ``pseudo-conversation,'' where the agent follows predetermined dialogue flows when interacting with children ~\cite{xuAreCurrentVoice2021}. This limitation could be especially problematic in creative storytelling activities, where children may generate a wide range of creative responses. \revision{Nevertheless, to overcome the technological limitations at the time, many studies employed Wizard-of-Oz approaches to explore the feasibility of children engaging in joint storytelling with a non-human partner, as perceived by the child. For example, Sun et al.~\cite{sunCollaborativeStorytellingRobot2017} had children aged 4 to 10 interacted with a robot secretly controlled by a researcher, periodically inserting new story content and relating it to the story created by the child. Children were found to meaningfully collaborate with the ``robot'' and enjoy their interactions. These Wizard of Oz studies suggested that it is possible to replicate the strategies utilized by human partners in a setting with other technological partners, thereby laying the foundation for our research}. With the advances in LLM, such human-led interactions can now be more feasibly simulated by AI agents. Thus, a primary objective of our current study is to utilize LLMs to emulate ``true conversation,'' which is defined as multiple parties 'mutually orienting to, and collaborating in order to achieve, orderly and meaningful communication'~\cite{enyiLanguageInteractionalDiscourse2015}

\subsection{Using Large Language Models for Child-facing Conversational Interfaces}
Pre-trained large language models (LLMs), such as GPT-3~\cite{brownLanguageModelsAre2020a} and GPT-4~\cite{GPT4}, have significantly advanced natural language processing (NLP) in recent years. 
These models, trained on colossal amounts of text data, can generalize to downstream tasks like text generation~\cite{weiEmergentAbilitiesLarge2022}. 
One key capability derived from the large model size is \textit{prompting}~\cite{weiEmergentAbilitiesLarge2022}, which allows individuals to provide specific textual instructions and examples~\cite{zamfirescu-pereiraWhyJohnnyCan2023} to guide the model's task execution. 
Although prompting LLMs may not consistently surpass benchmark models, it offers a lightweight approach to achieve competitive performance across a variety of tasks~\cite{chowdheryPaLMScalingLanguage2022,brownLanguageModelsAre2020a}. 
The concept of prompting LLMs remains a compelling research topic within the HCI community. 
Our work extends previous research by introducing a set of prompting techniques iteratively designed to optimize LLMs as story co-creators tailored for children’s educational purposes.\looseness=-1

CUIs powered by LLMs, such as ChatGPT~\cite{IntroducingChatGPT}, have demonstrated their potential in facilitating responsive and engaging dialogues. 
Unlike traditional CUIs, they are not constrained by a pre-determined dialogue paradigm or reliant on pre-written responses.
One of their standout features is the ability to generate contextually relevant and coherent responses based on the input they receive so far~\cite{weiEmergentAbilitiesLarge2022}. 
This capability positions them to achieve ``true conversation'', wherein they can effectively collaborate with children, ensuring a more natural and enriching dialogue~\cite{leeDAPIEInteractiveStepbyStep2023}.
Thus, this paper harnesses the adaptability and flexibility of LLMs, aiming to forge a dynamic and collaborative storytelling experience for children, thereby fostering their math language growth and stimulating their creative expression.

Recently, numerous researchers have endeavored to incorporate LLMs into educational technologies~\cite{kasneciChatGPTGoodOpportunities2023}. 
These adaptations have seen the utilization of LLMs in creating educational content~\cite{dijkstraReadingComprehensionQuiz2022a,gabajiwalaQuizMakerAutomatic2022a}, enhancing student engagement and interaction~\cite{baoCanHomeUse2019,taiImpactGoogleAssistant2020,abdelghaniGPT3drivenPedagogicalAgents2023}, and personalizing learning experiences~\cite{sailerAdaptiveFeedbackArtificial2023}. 
For instance, LLMs have been used to generate children's narratives~\cite{guanKnowledgeEnhancedPretrainingModel2020,alabdulkarimAutomaticStoryGeneration2021}, some of which have even been sold publicly~\cite{messickChatGPTAuthor1002023}. 
In a different research trajectory, several scholars have used LLMs to create intelligent learning partners capable of collaborating with humans~\cite{jiSystematicReviewConversational2022}, providing feedback~\cite{jiaAllinoneMultitaskLearning2021} and encouraging students~\cite{elshazlyEffectsArtificialIntelligence2021,taiImpactGoogleAssistant2020}. 
One common application involves employing LLMs as a conversational partner in written or oral form, such as in the context of task-oriented dialogues that offer language practice opportunities~\cite{elshazlyEffectsArtificialIntelligence2021}.
Building on these advancements in LLMs, our work aims to apply LLMs in CUIs to enable child-AI collaborative storytelling for mathematical language learning.\looseness=-1


\section{The Development Process of \textsc{Mathemyths}}
Grounded in the strong evidence that children develop math language through storytelling, and considering the unique capabilities of LLMs, we engaged in a design process to conceptualize \TOOL. Our aim was to create a co-creative storytelling system tailored for children aged 4-8, introducing them to mathematical language through voice-based interactions. This initiative serves as a case study to explore the potential of LLMs as educational partners for children.\looseness=-1

At its core, \TOOL is designed to collaboratively engage in storytelling by alternating turns with a child. 
The role of \TOOL within this dynamic is three-fold: 1) generating prompts to elicit narrative contributions from the children, 2) continuing the story by utilizing mathematical language based on the children's input, and \revision{3) providing scaffolding when children are hesitant to engage or show confusions about the prompts or the story. }The interaction flow is presented in Fig.~\ref{fig:all_in_one}. 
The example conversation between a child and \TOOL can be found in Appendix~\ref{sec:example_conversation}.

Given that using LLMs, particularly through prompt engineering, to develop educational tools is a relatively new domain, our design process began by applying educational principles from the existing literature reviewed above, which focus on engaging children in joint storytelling and math talk. We then engaged in a series of experiments to optimize \TOOL's capabilities for child-friendly and math-focused dialogue. In the remainder of this section, we outline 1) our overall design principles for \TOOL, 2) our method to prompt engineer LLM to serve as the story co-creator, 3) our process to evaluate the prompt-engineered LLM's performance, and 4) the implementation of the \TOOL system. \looseness=-1

\begin{figure*}
\centering
\includegraphics[width=\linewidth]{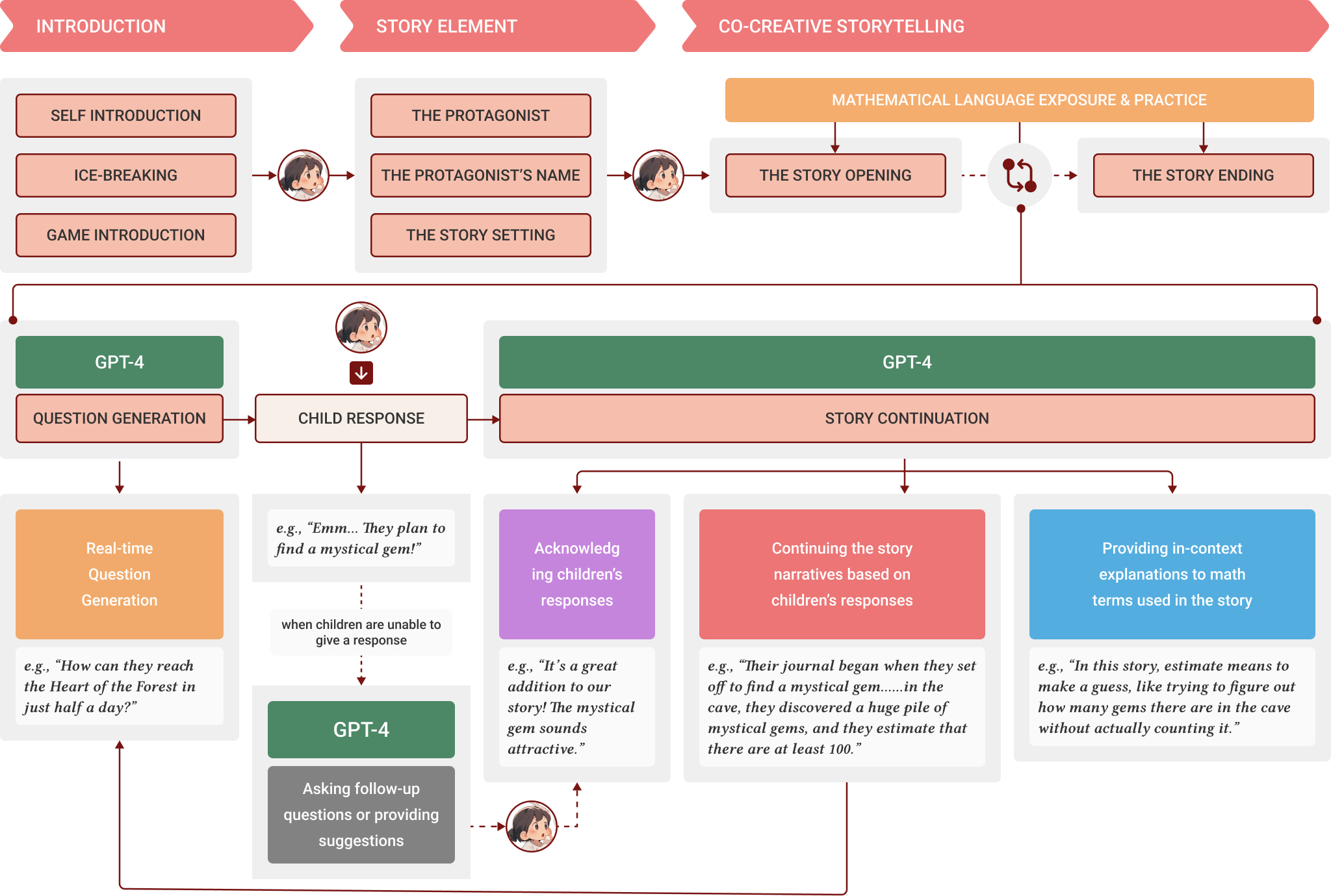}
\caption{\revision{The interaction flow of the \TOOL system.}}
\Description{The interaction flow of the Mathemyths system.}
\label{fig:all_in_one}
\end{figure*}

\subsection{Design Principles}
Drawing from the literature reviewed above, we incorporated two key design principles into the story co-creation interaction.
 
First, the conversation experience is designed to emulate a joint storytelling activity while the AI partner guides children in this process through using three-part dialogue moves that adhere to the \textit{question-feedback-scaffolding} framework~\cite{xuElinorTalkingMe2022}. Under this framework, \TOOL initiates the interaction with children by crafting an initial story plot based on children's input about key elements such as the protagonist and setting. It then poses open-ended \textbf{questions} to solicit how children wish the story should progress, such as ``\textit{What will Lucy and Nick do next to make sure they take an equal number of steps to reach the treasure?}'' After children's responses, \TOOL then offers \textbf{feedback} that includes both an acknowledgement of children's contribution (\eg ``Wonderful idea! That's a clever way to find the map.'') and a continuation the narrative in line with the children's input. \revision{In the cases when young children require additional support to generate a concrete idea, \TOOL provides \textbf{scaffolding} to encourage children to elaborate their ideas ~\cite{lamottBirdBirdInstructions1995, zhangStoryDrawerChildAI2022,zhangStoryDrawerCoCreativeAgent2021,sunCollaborativeStorytellingRobot2017}.
For instance, if the child provides a brief response without much detail, such as ``\textit{ask for help}'', the agent will ask a follow-up question, ``\textit{What a good point! Who should we ask for help? Can you tell me more about these helpful friends?}''
In other cases, if the child indicates confusions (e.g., \textit{`I don't know}'') or does not respond, \TOOL will scaffold the child by 
employing the ``hint and rephrase'' strategy to provide ideas for children to consider (e.g., ``\textit{Let's think together. Lucy and Nick can take turns counting their steps. How many steps do you think they should both count?}''). This strategy, simplifying the original open-ended question by suggesting a specific direction, is widely utilized in conversational agents to engage preschool-aged children ~\cite{graesserAutoTutor2012,graesserIntelligentTutoringSystems2012,xuElinorTalkingMe2022}.

To make the storytelling experience more suitable for young children, \TOOL creates story based on the quest archetype. This genre, prevalent in children's literature, focuses on a hero's journey to a particular place or to find an item, facing many challenges along the way ~\cite{huckChildrenLiteratureElementary2003}.  Language-wise, \TOOL is instructed to use simple, clear language that young children can easily understand, with sentences that are straightforward in structure. }\looseness=-1

Second, situated within the narrative structure, \TOOL is also designed to teach math language through exposure (i.e., using math terms in the plots generated by \TOOL) and practice (i.e., posing questions to encourage children to use math terms to continue the stories). \TOOL primarily focuses on six mathematical terms (\textit{sum}, \textit{estimate}, \textit{add}, \textit{subtract}, \textit{equal}, and \textit{half}). \revision{These terms and their related concepts align with the kindergarten and first grade Common Core State Standards~\cite{commoncorestatestandardsinitiativeMathematicsStandardsCommon2010,akkus2016common}. The children in these grades, usually aged between 5 and 7 years, fall in the middle of our target age range.} We also tested other terms from the Common Core Standard, such as ``divide'' and ``multiply'', but ultimately decided not to use those terms as they might be too complex for our younger participants. Children heard these terms used at least twice per story. \revision{Moreover, \TOOL is uniquely designed to not only introduce these terms but also provide explanations within the narrative context. For instance, when the term ``equal'' is used in a sentence like, ``Lucy and Nick take an equal number of steps,'' \TOOL explains it in a story-relevant way: ``This means Lucy and Nick take the same number of steps. If Lucy takes 5 steps, then Nick also takes 5 steps.'' These types of explanations, integrated within the context, are deemed more effective for young children than abstract definitions~\cite{xuElinorTalkingMe2022,dietzStoryCoderTeachingComputational2021,glaserNarrativebasedLearningPossible2009}. }

\subsection{Prompt Engineering}
The model underlying \TOOL was GPT-4\footnote{\url{https://openai.com/gpt-4}}, currently the most advanced LLM available for developers. Through prompt engineering, GPT-4 performed three primary tasks aligned with the question-feedback-scaffolding model: 1) generating questions to facilitate children's continuation of the story; 2) continuing the story based on children's responses, including in-context explanations of math terms; \revision{and 3) providing scaffolding with additional suggestions or follow-up questions, tailored to children's different responses.} 

The prompts were developed through an iterative process. We borrowed Brown et al.~\cite{brownLanguageModelsAre2020a} suggested prompt engineering strategies to draft our initial prompts. 
We then underwent a two-month-long revision process, which involved internal evaluations with the research team and evaluations with children, and this process led us to rapid iterations of the prompts.\looseness=-1

All these features were implemented using the OpenAI APIs\footnote{\url{https://platform.openai.com/docs/api-reference/chat}}.
The final version of prompts used in the user study is displayed in Appendix~\ref{sec:appendix_prompt} Table~\ref{tab:prompt_list}.

\subsubsection{Question Generation}
Following a similar prompt structure proposed in Brown et al.'s work~\cite{brownLanguageModelsAre2020a}, each prompt starts with a preamble which explains the prompt's purpose: ``\textit{Given an unfinished story, compose one single, straightforward question to prompt a 6-year-old to expand on the story.}''
The preamble is followed by a detailed list of instructions and multiple exemplars consisting of the input and the output. Specifically, we directed the model to pose questions concerning the main character's subsequent actions or emotions. These two story elements are not only important to the narrative's progression but are also tangible concepts that children can readily engage with and respond to ~\cite{paris2003assessing}. To enhance GPT-4 performance, we provided six exemplar questions generated by the research team. These questions were further reviewed and approved by two trained research assistants in the domains of children's verbal storytelling and mathematics education. 
The detailed prompt templates are shown in Appendix~\ref{sec:appendix_prompt} Table~\ref{tab:prompt_list}.

\subsubsection{Story Continuation} 
To facilitate the continuation of stories based on children's inputs, while integrating mathematical language instruction, we began by prompting the model to assume the role of a ``\textit{storytelling robot}\footnote{\revision{In the prompt, the model is designated to function as a ``storytelling robot''.
However, during interactions, we avoid describing our system as a robot to prevent potential confusion, given that it operates as a virtual agent.}}''. 
This prompt established the interactive framework of the LLM's responses.

The next part of the prompt was focused on context-setting. We supplied the model with a narrative scenario: ``\textit{Let's play a joint storytelling game where we build a story together with the main character being a [character] named [name] in the [setting].}'' The elements in brackets ([character], [name], [setting]) would be dynamically populated based on the responses provided by the children. 
\revision{We also set several high level rules to ensure the generated story content ``\textit{is simple and appropriate for a young child}'', adheres to the quest archetype,  and contains nothing unrelated to the story.}

For the model to produce coherent and math-focused responses, we articulated a specific sequence for it to follow. The model was tasked with acknowledging children's contributions, utilizing a minimum of two mathematical terms, and then offering in-context explanations. This was achieved via the following detailed prompt:
``\textit{First, acknowledge my addition and commend me in a short 10-word sentence. Next, continue the story by correctly using [mathematical term one] and [mathematical term two] or their variants within two distinct short sentences. Lastly, explain their meanings within a short sentence, grounding it firmly within the story's context.}''

During experimentation, we observed the LLM's tendency to occasionally overlook prior directives as interactions continued. To counter this behavior and maintain consistency, our specific instructions were reiterated with each turn. This ensured \TOOL's alignment with our intended conversational flow.


\subsubsection{Scaffolding} 
\revision{
In our study, we classified the situations where children require scaffolding into two main categories: 
(1) cases where they completely fail to continue the story, either by expressing uncertainty (e.g., saying ``\textit{I don't know}'') or by not responding;
(2) instances where they continue the story with only brief and vague responses, typically less than 5 words. 
To scaffold the first situation, our initial step is to determine whether the child is providing meta-comments (\eg ``\textit{I like this}'', ``\textit{please continue}'') or exhibiting signs of confusion (\eg ``\textit{I am not sure}''). 
This determination is made by prompting the LLM with a set of few-shot examples. 
These examples represent the most common types of responses observed in our rapid iteration tests with children. 
Following this identification, we utilize the model to simplify the question and offer hints related to the question, adhering to the ``hint and rephrase'' strategy.
In addressing the second situation, we monitor the child's responses. 
To scaffold the second situation, we employ the GPT-4 model to pose a follow-up question based on the child's preceding brief response. This approach involves first acknowledging what the child has said and then encouraging them to elaborate further on their contributions.
}

\subsection{Model Evaluation}
As part of the iterative design process, we conducted an evaluation to examine the technical capabilities of our LLM-powered system. This evaluation consisted of two components: first, assessing the model's ability to \textit{generate questions} to elicit children's responses to determine if these questions are fluent, inspirational, and logically relevant; second, evaluating the model's performance in \textit{continuing stories} that are logically relevant, engaging, and age-appropriate. 
Specifically, our goal was to confirm whether the prompt-engineered GPT-4 model performs satisfactorily compared to a human baseline in both tasks and whether its performance surpasses that of a fine-tuned GPT-3.5 model in the question generation task. \revision{ We chose the fine-tuned GPT-3.5 model as our baseline because it was the highest-performing model available that allowed fine-tuning during our development phase. Indeed, researchers have used a fine-tuned GPT-3.5 model in dialogic interactions with children, achieving better performance compared to other state-of-the-art models~\cite{leeDAPIEInteractiveStepbyStep2023}.}

All evaluations were carried out by three college student research assistants who had over a year of experience in educational research and extensive years of engagement in activities with children within the target age range of our study. These students underwent training until their inter-rater reliability reached a satisfactory level before beginning the actual evaluation.

\subsubsection{Evaluating Prompt Engineered GPT-4's Performance in Question Generation}

The generated questions are designed to inspire children to continue the story and encourage them to learn and use mathematical terms. In this evaluation, we compared our prompt-engineered GPT-4 model with a fine-tuned GPT-3 model. The fine-tuned GPT-3 model was trained\footnote{\url{https://platform.openai.com/docs/guides/fine-tuning}} using a subset of 100 out of 300 human-crafted questions developed by an author of the paper. Besides these two LLMs, we also included the other subset of human-crafted questions as our baseline.

The question evaluation included four matrices, namely \textit{readability}, \textit{inspiration}, \textit{story relevancy}, and \textit{math language relevancy}. We invited the three evaluators to score the questions across each of the four dimensions in a 5-point Likert scale. The inter-rater reliability among these evaluators was satisfactory with an average of intra-class correlation of 0.82 across four dimensions. Details of these matrices are presented in Table~\ref{tab:expert_evaluation_question}.

\begin{table*}
\small
\centering
\caption{Metrics used in evaluating prompt engineered GPT-4's performance in question generation. The average intraclass correlation coefficient (\textit{ICC}) scores are displayed.}
\label{tab:expert_evaluation_question}
\begin{tabularx}{\textwidth}{lXl}
\toprule
\textbf{Metric} & \textbf{Description} & \textbf{\textit{ICC}}\\
\hline
Readability & The generated question is written in understandable English, using proper grammar and vocabulary. & 0.758 \\
Inspiration & The generated question stimulates children's thinking and provokes them to articulate their thoughts. & 0.870 \\
Story Relevancy & The generated question aligns with the narrative plotline. & 0.751\\
Math Language Relevancy & The generated question requires children's understanding or use of specific mathematical terms. & 0.907\\
\bottomrule
\end{tabularx}
\end{table*}


\begin{figure*}
\centering
\includegraphics[width=\linewidth]{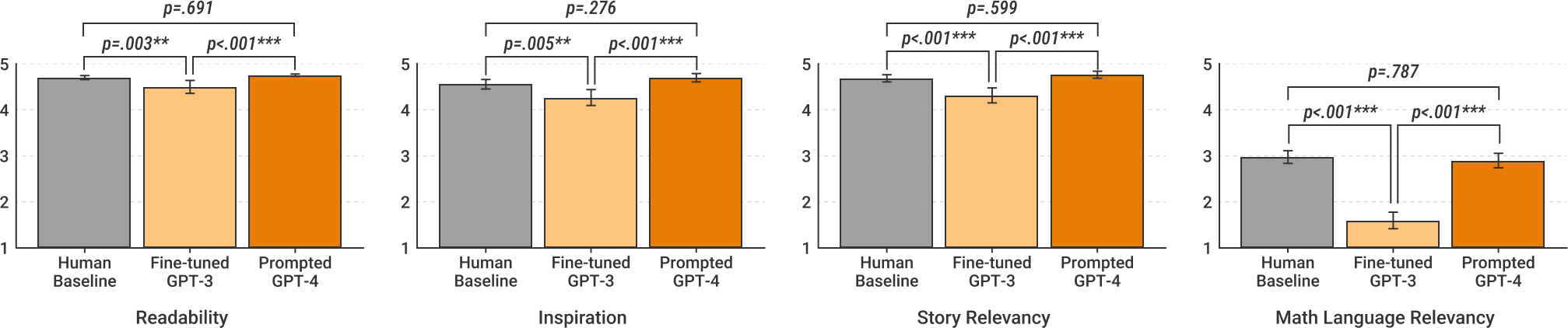}
\caption{\correct{Bar plots illustrating the distribution of data and the results from the ANOVA post-hoc Tukey's HSD test regarding the question generation evaluation. Statistically significant results are reported as \(p < 0.05^{*}\), \(p < 0.01^{**}\), \(p < 0.001^{***}\). Error bars represent 95\% confidence intervals (CIs).}}
\Description{Bar plots illustrating the distribution of data and the results from the ANOVA post-hoc Tukey's HSD test regarding the question generation evaluation.}
\label{fig:question_generation}
\end{figure*}

An ANOVA was conducted to compare the ratings of questions generated by human, fine-tuned GPT-3.5, and our prompt-engineered GPT-4 model. 
The results indicated significant differences among the sources for all four dimensions (Readability: \(F(2) = 9.415, p < .001^{***}\),  Inspiration: \(F(2) = 11.357, p < .001^{***}\), Story Relevancy: \(F(2) = 17.327, p < .001^{***}\), Math Language Relevancy: \(F(2) = 90.687, p < .001^{***}\)).
A post hoc Tukey's HSD test further revealed that questions generated by both humans and GPT-4 exhibited higher levels of readability, inspiration, and relevance to the story's plot and mathematical language when compared to questions generated by the fine-tuned GPT-3.5 model (Fig.~\ref{fig:question_generation}).
The evaluation ratings between human-generated and GPT-4 generated questions were comparable, further suggesting the feasibility and promise of GPT-4 models in this context.
\revision{However, the questions, either generated by humans or GPT-4, were scored relatively lower in the relevance of mathematical language as compared to other dimensions \correct{but still higher than GPT-3.5}. This arises from the complexity of creating questions that both meaningfully advance the storyline and focus on specific math terms. To preserve the story's continuity, the questions' relevance to stories was often prioritized. Despite these, children should still encounter a sufficient number of questions aimed at enhancing their understanding and use of mathematical concepts.}
\looseness=-1

\subsubsection{Evaluating Prompt Engineered GPT-4's Performance in Story Continuation}
\label{sec:model_evaluation_story}

\TOOL is designed to continue the story adaptively based on children's contributions and integrate mathematical language within the story plot. 
To evaluate whether our prompt-engineered GPT-4 model achieved these two goals, we compared the GPT-4 generated stories with those generated by humans as our baseline, given that other existing models were not tailored to continue stories but rather generate new stories. 
GPT-4 generated stories were created using a self-chat technique ~\cite{ghandehariounApproximatingInteractiveHuman2019}, which enabled our dialogue system to engage in self-generated conversations, mimicking a child's interactions, and collaboratively creating stories. 
Human-generated stories were based on the same instructions we provided to prompt GPT-4. The process for generating the dataset is detailed in Appendix~\ref{sec:appendix_dataset}. 
Subsequently, we presented the three evaluators with pairs of stories, one generated by GPT-4 and the other by humans. 
\revision{Based on a widely recognized method for evaluating AI-generated dialogues~\cite{leeDAPIEInteractiveStepbyStep2023,liACUTEEVALImprovedDialogue2019}, we asked them to make a choice from each pair based on four dimensions: \textit{preference}, \textit{perceived creativity}, \textit{mathematics relevancy}, and \textit{readability}.} 
Table ~\ref{tab:expert_evaluation_story} provides a detailed explanation of these metrics.

\begin{table*}
\small
\centering
\caption{Metrics used in evaluating prompt engineered GPT-4's performance in story continuation.}
\label{tab:expert_evaluation_story}
\begin{tabular}{ll}
\toprule
\textbf{Metric} & \textbf{Description}\\
\hline
Preference & Who would you prefer to collaborate with to create a story? \\
Perceived Creativity & Whose story sounds more creative and contains more twists and turns?\\
Mathematics Relevancy & Whose story contains more mathematical elements?\\
\revision{Readability} & \revision{Whose story is more easily comprehended by children in our target age range?}\\
\bottomrule
\end{tabular}
\end{table*}

\begin{figure*}
\centering
\includegraphics[width=0.8\linewidth]{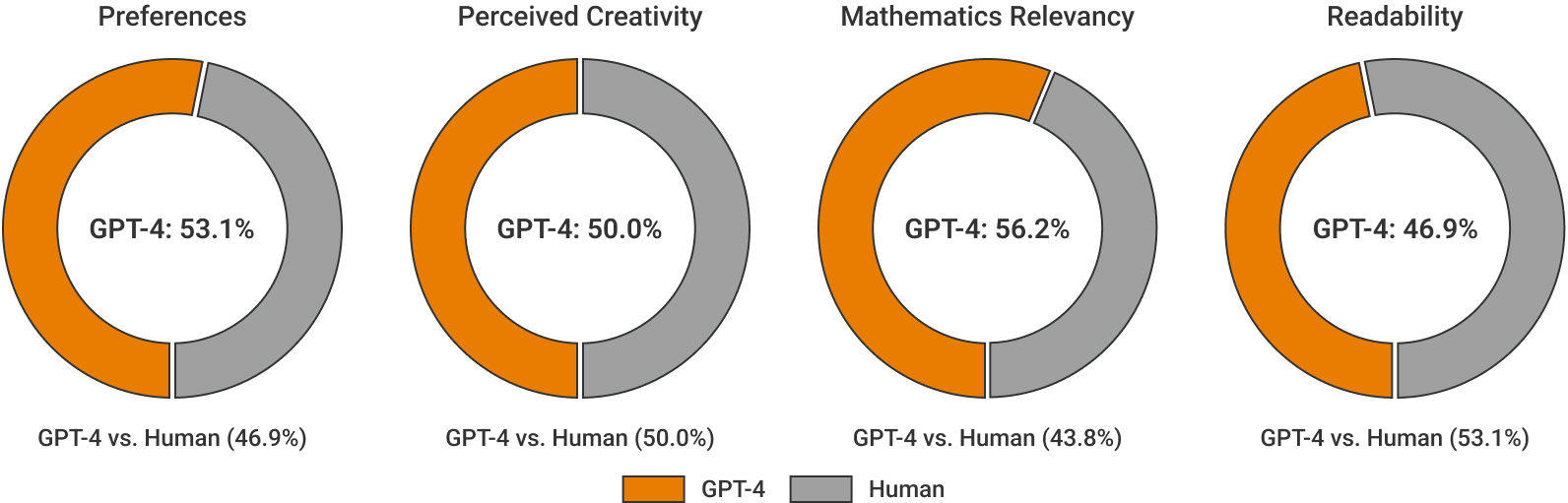}
\caption{\correct{The evaluation results on four metrics of story continuation.}}
\Description{The evaluation results on four metrics of story continuation.}
\label{fig:story_preference}
\end{figure*}

\revision{The results of this evaluation are displayed in Fig.~\ref{fig:story_preference}. Overall, our evaluators indicated that the story pairs, where one was generated by humans and the other by GPT-4, were quite comparable in quality across all dimensions, particularly in perceived creativity (50.0\% vs. 50.0\%). Notably, there was a marginal preference for stories generated by AI (53.1\% vs. 46.9\%), with these stories also being perceived as integrating more mathematical elements (56.2\% vs. 43.8\%).  
GPT-4's advantages in relating stories to mathematical elements are likely due to the fact that this task capitalizes on LLMs' capabilities of creating seemingly logical connections of language elements based on specific instructions.
However, human-generated stories were slightly favored for their readability (46.9\% vs. 53.1\%). A follow-up discussion with the evaluators revealed that the slightly lower readability in AI-generated stories primarily stemmed from instances where GPT-4 produced text that, while structurally coherent in its language, occasionally lacked logical consistency. This observation aligns with existing research underscoring the inherent limitations ~\cite{liHaluEvalLargeScaleHallucination2023,jiSurveyHallucinationNatural2023}.}



\subsection{System Implementation}
\TOOL is a desktop application designed to run on a MacBook computer with built-in microphones and speakers. 
This application, written in Python, does not rely on any graphical interface for interaction. 
Instead, users can navigate \TOOL simply by voice. 
Upon launch, users have the flexibility to specify both the number of dialogue rounds and the mathematical terms they wish to learn. 
For voice-based interaction, \TOOL leverages the Google Cloud Speech-to-Text API\footnote{\url{https://cloud.google.com/speech-to-text}} to recognize users' verbal inputs. 
Additionally, the Elevenlabs Speech Synthesis service\footnote{\url{https://beta.elevenlabs.io/speech-synthesis}} is employed to generate realistic speech for the agent's responses.



\section{User Study}
\label{sec:user_study}
To understand how \TOOL might support children's engagement and math language learning, we conducted a user study where children were randomly assigned to either co-create stories with \TOOL or with a friendly and skilled human partner. 
We examined children's gains in mathematical language knowledge, their engagement, creation performance, and enjoyment in the storytelling activities, and \TOOL's performance.


\subsection{Procedure}
Participants were recruited from two public libraries in a mid-west city in the U.S. throughout the summer of 2023. This study was approved by the Institutional Review Board of the authors' institution. After obtaining parental consent and child assent, participants were introduced to a research assistant and began the three-part study procedure which consisted of a baseline pretest of children's math language, the story co-creation activity, and  a post-test assessing math language, along with a survey with the children. Details of the math language pre- and post-test questionnaire are described in the Evaluation Metrics section below. 


\begin{figure}
\centering
\includegraphics[width=\linewidth]{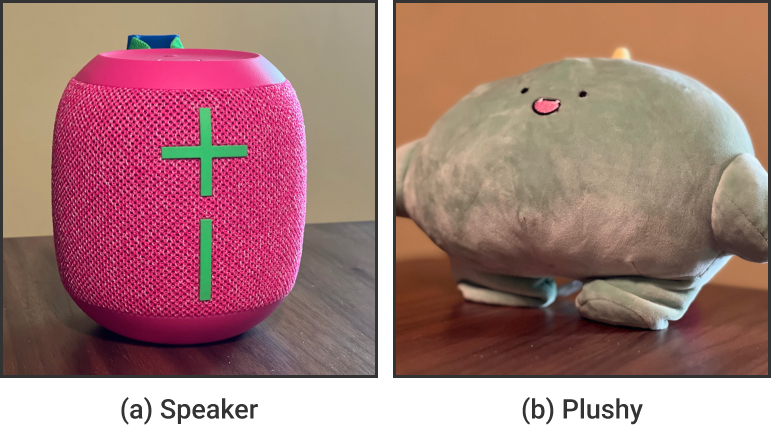}
\caption{The (a) plushy and (b) speaker used in our user study.}
\Description{The (a) plushy and (b) speaker used in our user study.}
\label{fig:study_setting}
\end{figure}

For the story co-creation activity, children were randomly assigned to partner either with a human (i.e., a train research assistant) or AI (i.e., \TOOL) presented as a colorful speaker as displayed in Fig.~\ref{fig:study_setting}(a), for co-creating two short stories. 
In both conditions, children participated in a warm-up session to familiarize themselves with the story co-creation activity. This session also served to illustrate how the AI or human partner would precisely respond to their interactions, including scenarios where the children asked questions or chose to remain silent. 
To facilitate children's story creation, they were handed a plushy (Fig.~\ref{fig:study_setting}(b)) in case they wished to act out the story or just hold on to it. 
After the warm-up session, children were free to interact with the AI without interference or minimal redirection and instructions from the researcher. 
For both \TOOL and human conditions, the creation of each story typically lasted 6 minutes, resulting in a total duration of approximately 12 minutes for each child. \looseness=-1

All parts of the procedure were video recorded for later transcription of the stories and coding of open-ended questions. Participants had the option of completing all three parts in a single session or completing the baseline assessment one day and returning another day to complete the story creation activity and learning evaluation. While children were completing the story creation activity, parents were provided with a questionnaire on an iPad to collect demographic information about their child. Once the child completed the procedure they were given a small prize and a book as a token of appreciation and parents received a twenty-dollar Visa gift card for their participation.

\subsection{Evaluation Metrics}
\label{sec:evaluation_metrics}
In this user study, our focus was on children's learning of target math vocabulary, the quality of the stories they created, as well as their engagement and enjoyment throughout the process. \revision{For all measures requiring children's responses, the questions were orally narrated and asked by an experimenter, and the children were expected to answer orally as well. Thus, children did not need reading or writing skills to answer those questions.}

\subsubsection{Learning of Math Language}
\label{sec:math_language_assessment}
The research team developed a 24-item questionnaire focusing on the six math terms (\ie \textit{sum}, \textit{estimate}, \textit{add}, \textit{subtract}, \textit{equal}, \textit{half}), based on the Common Core State Standards~\cite{akkus2016common,commoncorestatestandardsinitiativeMathematicsStandardsCommon2010} and Purpura et al. ~\cite{purpuraNonlinearRelationsApproximate2015}. These terms were assessed through four dimensions:  \textit{definition}, \textit{recall}, \textit{transfer}, and \textit{practice}, with each dimension comprising six questions, one question for each target term. For the \textit{definition} dimension, children were prompted to provide the definition of each of the six math terms. In the \textit{recall} dimension, they were asked to identify each term when its definition was given. For \textit{transfer} questions, children needed to provide an appropriate math term based on a provided real-life scenario. Lastly, in the \textit{practice} dimension, children were asked to use the target math terms to freely generate descriptions for a picture featuring fruits of various types, amounts, sizes, and colors. The complete questionnaire can be found in Appendix~
\ref{sec:appendix_math}.\looseness=-1

\revision{During the pre and post-tests, we utilized the same set of questionnaires with slight modifications. For example, in the transfer set of questions, nouns were changed – (\ie ``you have a jar filled with candies'' became ``you have a net filled with butterflies''). Additionally, all items were presented in a random order. These modifications were implemented to reduce the likelihood that children were merely replicating their answers from the baseline assessment.}

To calculate the scores from this questionnaire, the \textit{definition} and \textit{practice} items, being open-ended, were rated on a binary scale: 0 for incorrect and 1 for correct responses. For the \textit{recall} and \textit{transfer} items, children first attempted to answer through free recall. If unsuccessful, they were then provided with three options to choose from. Scores were assigned as follows: 2 for correct free recall answers, 1 for correct answers chosen from the given options, and 0 for incorrect answers. Based on this scoring system, we calculated a total score by summing the points across all items, with a possible range from 0 to 36. 
The Cronbach's alpha of these items was 0.92.

\subsubsection{Engagement}
To understand children's engagement in the story co-creation activity with either a human or AI partner, we analyzed their responses using three key indicators: whether they provided a verbal response, the length of each response, and the nature of their responses. The nature of responses was categorized into four types: meta-comments reflecting the child's thought process (e.g., ``I think so.''), expressions of uncertainty (e.g., ``I don't know''), brief responses with fewer than five words lacking sufficient details, and substantial responses containing meaningful details that advance the story. Two trained research assistants were responsible for the coding. The first coded transcripts for all participants, and the second coded 30\% of the data for quality control purposes. Between these two coders, there was a 100\% agreement rate across all items.\looseness=-1

\subsubsection{Story Creation Performance}
We analyzed the children's performance on their story creation using two indicators: the number of ideas and elaborations, based on prior research ~\cite{wongSpokenNarrativeAssessment2016,zhangStoryDrawerChildAI2022},
An idea is defined as a character and its associated action within a single utterance. For example, ``The cat climbs the tree'' is one idea, with ``cat'' as the character and ``climbs'' as the action. An elaboration refers to additional details provided to enrich the idea, such as when, where, and why. For instance, ``The cat climbs the tree because it's curious and loves to explore'' adds reasoning for the character's action. One researcher coded the transcripts for all participants, and another coded 30\% of the participants' transcripts as a quality check. This process resulted in a satisfactory Intraclass Correlation Coefficient of 0.85.
\looseness=-1

\subsubsection{Enjoyment}
To measure children's enjoyment of interacting with AI or humans, we adapted a 4-item questionnaire based on Waytz et al.'s work ~\cite{waytzMindMachineAnthropomorphism2014}. The questions were: 
(1) \textit{Were you happy when you told stories with []?}; 
(2) \textit{Did you feel comfortable telling stories with []?}; 
(3) \textit{Would you like to have a reading partner like []?}; and 
(4) \textit{Do you want to tell another story with []?}. 
\revision{The questionnaire asked children to indicate their level of agreement with each item, using a pictorial scale to facilitate their responses. For example, in response to the first question, children were presented with five options of ``really happy,'' ``happy,'' ``kind of happy,'' ``a little bit happy,'' and ``not happy at all.'' Each option was associated with a circle varying in size from large to small, with the researcher pointing to each circle while describing the corresponding option. The size of the circle corresponded to the level of happiness, with the largest circle representing the most happiness and the smallest indicating the least. To ensure that children based their responses on \TOOL and not on other voice assistants they may have used at home, we reminded them at the beginning that we were specifically asking about the AI they had just created a story with, while also pointing at the device. The Cronbach's alpha internal consistency is acceptable at 0.77. }\looseness=-1

\subsection{Participants}

\begin{table*}
\small
\centering
\caption{Participant information by study conditions. \correct{An independent t-test and a series of Chi-squared tests revealed that there are no significant differences between the two conditions in terms of baseline math language skills, age, children's gender, race/ethnicity, caregiver education, or prior use of conversational agents.}}
\label{tab:participant_information}
\begin{tabularx}{0.77\textwidth}{lcccc}
\toprule
 & \textbf{Full Sample} & \textbf{AI} & \textbf{Human} & \textbf{Difference}\\
\hline
Math Language Baseline & 16.686 & 17.105 & 16.188 & \(T(33) = 0.279, p = .391\) \\
\textit{Age} & & & & \(\chi^2(1) = 0.308, p = .579\) \\
\ \ \ \ \ \revision{Full sample} & \revision{6.000 (\(N = 35\))} & \revision{6.105 (\(N = 19\))} & \revision{5.875 (\(N = 16\))} & \\
\ \ \ \ \ \revision{4-5-year-olds} & \revision{4.462 (\(N = 13\))} & \revision{4.333 (\(N = 6\))} & \revision{4.571 (\(N = 7\))} & \\
\ \ \ \ \ \revision{6-8-year-olds} & \revision{6.909 (\(N = 22\))} & \revision{6.923 (\(N = 13\))} & \revision{6.889 (\(N = 9\))} & \\
\textit{Gender} & & & & \(\chi^2(1) = 0.046, p = .830\) \\
\ \ \ \ \ Female & 54.29\% (\(N = 19\)) & 52.63\% (\(N = 10\)) & 56.25\% (\(N = 9\)) & \\
\ \ \ \ \ \revision{Male} & \revision{45.71\% (\(N = 16\))} & \revision{46.37\% (\(N = 9\))} & \revision{43.75\% (\(N = 7\))} & \\
\textit{Race/Ethnicity} & & & & \(\chi^2(4) = 4.500, p = .343\) \\
\ \ \ \ \ Black & 11.43\% (\(N = 4\)) & 15.79\% (\(N = 3\)) & 6.25\% (\(N = 1\)) & \\
\ \ \ \ \ Latino & 8.57\% (\(N = 3\)) & 5.26\% (\(N = 1\)) & 12.50\% (\(N = 2\)) & \\
\ \ \ \ \ White & 60.00\% (\(N = 21\)) & 68.42\% (\(N = 13\)) & 50.00\% (\(N = 8\)) & \\
\ \ \ \ \ Others & 17.14\% (\(N = 6\)) & 10.53\% (\(N = 2\)) & 31.25\% (\(N = 5\)) & \\
\textit{Parents' Education} & & & & \(\chi^2(2) = 3.240, p = .198\) \\
\ \ \ \ \ Above Bachelor's degree & 34.29\% (\(N = 12\)) & 47.37\% (\(N = 9\)) & 20.00\% (\(N = 3\)) & \\
\ \ \ \ \ Bachelor's degree & 54.29\% (\(N = 19\)) & 42.11\% (\(N = 8\)) & 68.75\% (\(N = 11\)) & \\
\ \ \ \ \ Less than Bachelor's degree & 11.43\% (\(N = 4\)) & 
10.53\% (\(N = 2\)) & 12.50\% (\(N = 2\)) & \\
\textit{Usage of CA} & & & & \(\chi^2(4) = 1.689, p = .793\) \\
\ \ \ \ \ Daily & 17.14\% (\(N = 6\)) & 21.05\% (\(N = 4\)) & 12.50\% (\(N = 2\)) & \\
\ \ \ \ \ Weekly & 17.14\% (\(N = 6\)) & 21.05\% (\(N = 4\)) & 12.50\% (\(N = 2\)) & \\

\ \ \ \ \ Monthly & 14.29\% (\(N = 5\)) & 15.79\% (\(N = 3\)) & 12.50\% (\(N = 2\)) & \\
\ \ \ \ \ Less than once a month & 28.57\% (\(N = 10\)) & 21.05\% (\(N = 4\)) & 37.50\% (\(N = 6\)) & \\
\ \ \ \ \ Never & 22.86\% (\(N = 8\)) & 21.05\% (\(N = 4\)) & 25.00\% (\(N = 4\)) & \\
\hline
\ \ \ \ \ N & 35 & 19 & 16 & \\
\bottomrule
\end{tabularx}
\end{table*}

We recruited 35 participants (19 female, 16 male) aged between four and eight (\(M = 6\), \(SD = 1.35\)) by advertising in local libraries, childcare centers, and through snowball sampling. 
Participants were randomly allocated to either the experimental condition using our system (\(N = 19\)) or the control condition involving a human partner (\(N = 16\)). 
All the participants' predominant home language is English. 
No significant differences were observed between the two conditions in terms of children's gender, age, baseline math language skills, race/ethnicity, prior use of conversational agents, or caregiver education (Table~\ref{tab:participant_information}).


\subsection{Results}
In this section, we discuss our findings regarding children's learning of math language, their engagement, performance in story creation, as well as enjoyment with either a human partner or AI, \TOOL. \revision{It's important to note that our participants ranged in age from four to eight years old. This provided a valuable opportunity to examine how children at different developmental stages interact with AI. Thus, we divided the child participants into two age groups: a younger group (ages 4 to 5) and an older group (ages 6 to 8) ~\cite{dietzStoryCoderTeachingComputational2021}. This division allowed us to investigate whether and how developmental differences might influence our results. }

\begin{figure*}
\centering
\includegraphics[width=0.5\linewidth]{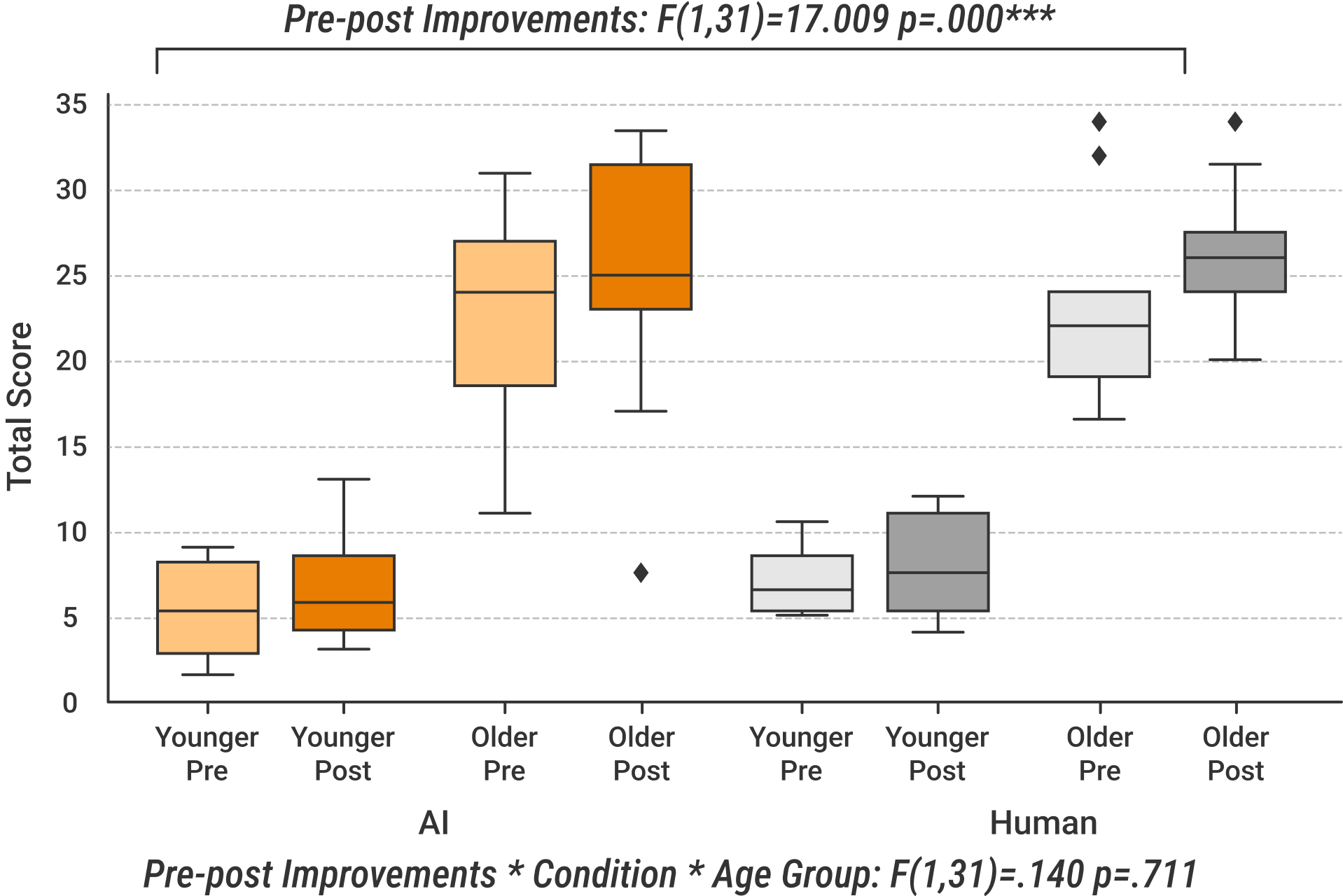}
\caption{\correct{Box plots illustrating the data distribution and the results of \revision{a two-way repeated-measures mixed ANOVA for the pre-post-test, using condition and age group as covariates,} in the mathematical language assessment. Statistically significant results are reported as \(p < 0.05^{*}\), \(p < 0.01^{**}\), \(p < 0.001^{***}\).}}
\Description{Box plots illustrating the data distribution and the results of a two-way repeated-measures mixed ANOVA for the pre-post-test, using condition and age group as covariates, in the mathematical language assessment.}
\label{fig:mathematical_language_total}
\end{figure*}

\subsubsection{Learning of Math Language}
\revision{Our analysis focused on examining if there was an improvement in children's understanding of targeted math terms from the pretest to the posttest, first considering the effects by condition (i.e., AI versus human) and then examining if these gains varied between the younger and older age groups. To address these questions, we carried out a two-way repeated-measures mixed ANOVA, using condition and age group as covariates.  
}

\revision{As shown in Figure~\ref{fig:mathematical_language_total}, the results indicated significant improvements between the pre- \correct{(\(M = 16.686\), \(SD = 9.578\))} and post-test \correct{(\(M = 18.971\), \(SD = 1.526\))} math language total scores \correct{(\(F(1,31) = 17.009, p < .000^{***}\))}. Moreover, these learning gains were found to be comparable across the AI \correct{(pre: \(M = 17.105\), \(SD = 9.882\); post: \(M = 19.500\), \(SD = 11.074\))} and human \correct{(pre: \(M = 16.187\), \(SD = 9.501\); post: \(M = 18.344\), \(SD = 10.158\))} conditions \correct{(\(F(1,31) = 0.012, p = .912\))}. Statistically, children’s learning gains showed \correct{marginal} variation across different age groups \correct{(\(F(1,31) = 3.219, p = .083\))}. On a descriptive level, older children seemed to benefit more from the co-creation activity than younger children, regardless of their interaction with \TOOL or a human partner. Thus, this age difference did not seem to stem from the nature of interacting with AI, but rather might be due to the fact that older children already have a better understanding of the underlying math concepts in the first place, which might amplify their comprehension of the associated math language.}

\revision{To further unpack children's learning gains across each dimension (i.e., definition, recall, transfer, practice), we carried out additional repeated-measures ANOVA analyses. These analyses used experimental conditions and age group as covariates, following the same model we applied in analyzing the total scores. Overall, we found that children's gain in math language appeared to be primarily driven by their improved scores in answering transfer and practice questions, as the pre-to-post-test gains were statistically significant in these dimensions, but not in definition and recall. However, when analyzing the pre-to-post gains between the story creation partners (human vs AI), AI partners showed a notable advantage in enhancing children's performance in definition questions \correct{(\(F(1,18) = 8.308, p = .010^{**}\); pre: \(M = 2.737\), \(SD = 1.939\); post: \(M = 3.368\), \(SD = 2.047\))}, while children with human partners did not show significant improvement in this dimension \correct{(\(F(1,15) = 0.015, p = .903\); pre: \(M = 2.813\), \(SD = 1.621\); post: \(M = 2.844\), \(SD = 1.777\))}. Further analysis by age group revealed a consistent pattern across all dimensions: older children showed greater improvements than younger ones, although these differences were not statistically significant.
}


\revision{Taking together,  \TOOL has shown to be as effective as a human partner in supporting children's math language learning through storytelling, while also displaying a stronger advantage in aiding definition comprehension. Moreover, children from both younger and older age groups benefited from this activity, whether interacting with an AI or human, though older children showed a slightly greater advantage.}

\looseness=-1


\subsubsection{Engagement}

\begin{figure*}
\centering
\includegraphics[width=\linewidth]{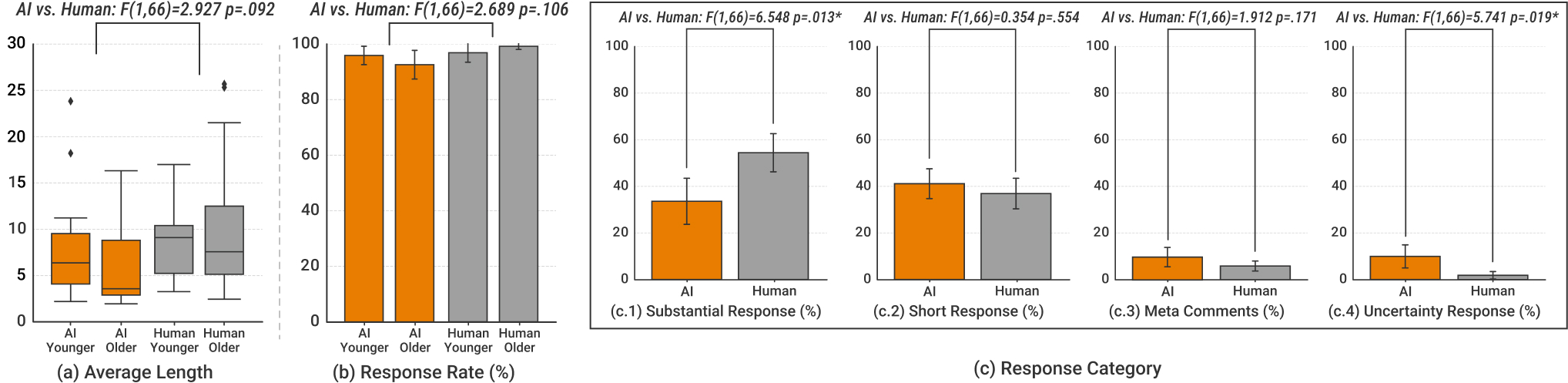}
\caption{\revision{Box plots and bar plots illustrating the distribution of data and the results from the two-way ANOVA regarding children's verbal engagement. Statistically significant results are reported as \(p < 0.05^{*}\), \(p < 0.01^{**}\), \(p < 0.001^{***}\). Error bars represent 95\% confidence intervals (CIs).}}
\Description{Box plots and bar plots illustrating the distribution of data and the results from the two-way ANOVA regarding children's verbal engagement.}
\label{fig:engagement}
\end{figure*} 

In our study, children actively engaged in the story co-creation activity with \TOOL. Focusing on the quantity of their engagement (i.e., response rate and length), children responded to over 95\% of the questions posed by \TOOL, with their responses averaging between 6-10 words in length. Generally, children tended to be more responsive and provided longer answers to questions asked by the human partner. \revision{Notably, the observed differences in responsiveness and response length between interactions with AI and humans appeared to be predominantly exhibited by older children. 
As shown in Figure~\ref{fig:engagement}, younger children did not differentiate in their behavior towards the AI versus a human: in both conditions, younger children exhibited almost exactly the same rate (AI: \correct{\(M = 96.002\), \(SD = 5.860\)}; Human: \correct{\(M = 96.922\), \(SD = 6.511\)}) and average response length (AI: \correct{\(M = 8.414\), \(SD = 6.579\)}; Human: \correct{\(M = 8.968\), \(SD = 4.328\)}). However, older children displayed a different pattern of interaction with humans, characterized by more frequent (AI: \correct{\(M = 92.668\), \(SD = 13.436\)}; Human: \correct{\(M = 99.246\), \(SD = 2.456\)}) and lengthier (AI: \correct{\(M = 6.064\), \(SD = 4.546\)}; Human: \correct{\(M = 10.358\), \(SD = 7.267\)}) responses compared to their interactions with the AI. However, this observed pattern did not reach statistical significance: The results of two-way ANOVA analyses indicated that there were no significant differences between younger and older participants across all engagement dimensions. 
}

In terms of the nature of children's responses, the majority were actual answers that advanced the stories, either substantial or short. This trend was consistent, regardless of whether children partnered with \TOOL or a human. However, we observed that children were less likely to provide substantial responses when interacting with \TOOL compared to a human. Additionally, interactions with \TOOL elicited a higher incidence of uncertain responses, like ``I don't know'', though still relatively infrequent, at a rate of 9.9\%, as opposed to 1.8\% when interacting with humans. This increased frequency of uncertainty in the AI condition might be attributed to \TOOL's tendency to pose unusual or fantasy-oriented questions. For instance, \TOOL asked questions like how the characters can use a cloud to speed up their journey, which are less realistic than questions asked by a researcher such as ``How can they get through the traffic circle faster?'' Nevertheless, the use of unusual and imaginative elements in responses may not necessarily be negative, and is actually a common feature in children's literature, especially for younger audiences. \revision{Our analysis of uncertain responses by younger versus older children revealed an interesting trend. While children in general exhibited a higher rate of uncertainty when responding to AI, this tendency was more pronounced among older children. Specifically, younger children showed a 7.6\% rate of uncertain responses to AI, compared to 10.9\% for older children. In contrast, with human interaction, the rate of uncertain responses was only 3.5\% for younger children and even lower at 0.5\% for older children.}

\looseness=-1

\subsubsection{Story Creation Performance}

\begin{figure*}
\centering
\includegraphics[width=0.6\linewidth]{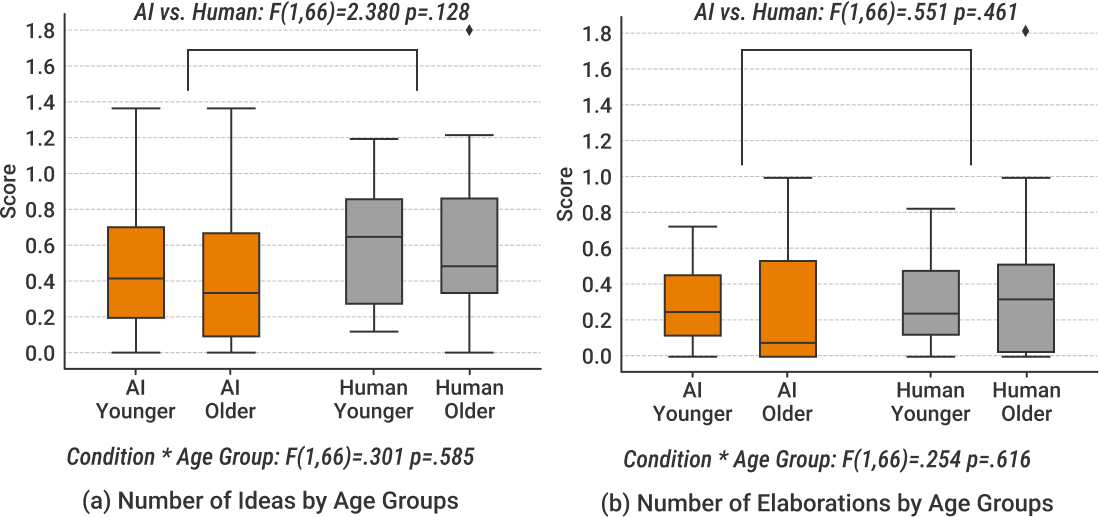}
\caption{\revision{Box plots illustrating the data distribution, along with the results of the two-way ANOVA comparing the story creation performance of children, using condition and age group as covariates.} Statistically significant results are reported as \(p < 0.05^{*}\), \(p < 0.01^{**}\), \(p < 0.001^{***}\).}
\Description{Box plots illustrating the data distribution, along with the results of the two-way ANOVA comparing the story creation performance of children, using condition and age group as covariates.}
\label{fig:creation_performance}
\end{figure*}

We then focused on the quality of the stories children created, measured by the number of ideas and elaborations. On average, a child produced about 0.5 idea per prompt during interactions with either AI or humans (AI: \correct{\(M = .443\), \(SD = .392\)}; Human: \correct{\(M = .616\), \(SD = .404\)}), amounting to nearly one idea every two prompts. In terms of elaboration, children typically offered 0.3 detailed elaboration per turn, equating to approximately 1 elaborate response every three prompts interactions with both AI and humans (AI: \correct{\(M = .285\), \(SD = .241\)}; Human: \correct{\(M = .360\), \(SD = .390\)}). \revision{As shown in Figure~\ref{fig:creation_performance}, a two-way ANOVA, examining experimental conditions and age groups, indicated that there were no significant differences in the quality of story creation between the \TOOL and human conditions \correct{(ideas: \(F(1,66) = 2.380, p = .128\); elaboration: \(F(1,66) = .551, p = .461\))}, nor between younger and older children \correct{(ideas: \(F(1,66) = .165, p = .686\); elaboration: \(F(1,66) = .130, p = .720\))}. 
}

\revision{Nevertheless, it might be quite surprising that younger children performed on par with their older peers in this metric, especially for those who interacted with the \TOOL (Figure~\ref{fig:creation_performance}), despite younger children's presumably less developed language ability. Yet, this might be due to the scaffolding features that we included, such as posing follow-up questions and providing hints when children did not formulate substantive responses. These features appeared to be effective in further eliciting children's responses and engaging them in back-and-forth interactions around one prompt. Thus, the scaffolding features could have reduced the presumed differences in the extent of idea elaboration between older and younger children.}

\subsubsection{Enjoyment}
\begin{figure*}
\centering
\includegraphics[width=\linewidth]{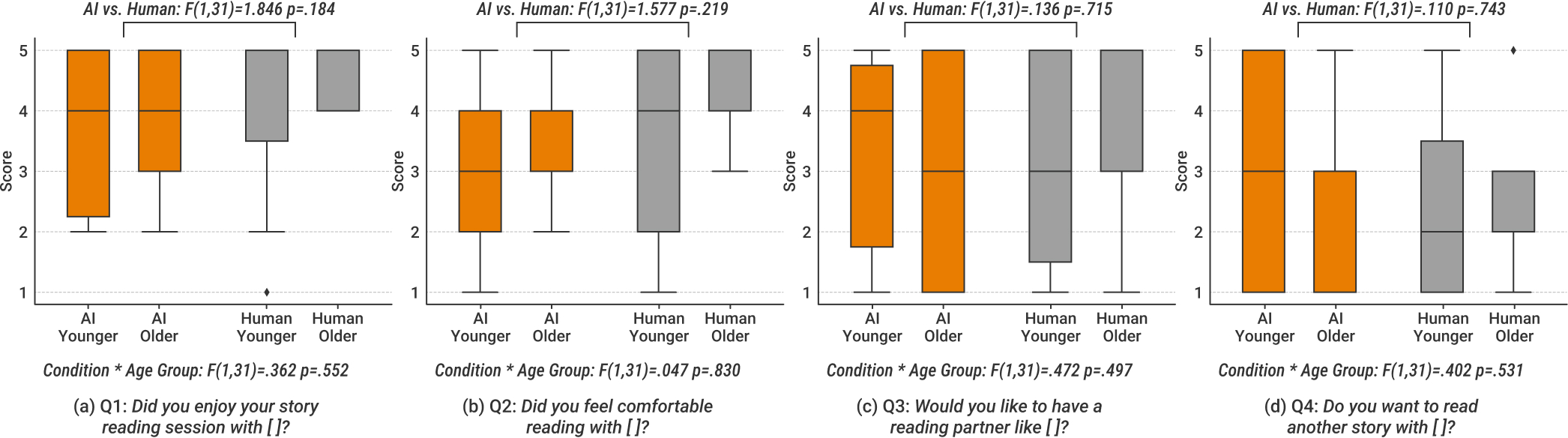}
\caption{\revision{Box plots illustrating the data distribution, along with the results of the two-way ANOVA comparing children's enjoyment, using condition and age group as covariates. Statistically significant results are reported as \(p < 0.05^{*}\), \(p < 0.01^{**}\), \(p < 0.001^{***}\).}}
\Description{Box plots illustrating the data distribution, along with the results of the two-way ANOVA comparing children's enjoyment, using condition and age group as covariates.}
\label{fig:enjoyment}
\end{figure*}

As shown in Figure~\ref{fig:enjoyment}, children's responses to the survey items indicated an overall positive perception of the story co-creation experience. Descriptively, they showed a more favorable reaction to reading with a human partner in terms of enjoyment (AI: \correct{\(M = 3.68\), \(SD = 1.250\)}; Human: \correct{\(M = 4.31\), \(SD = 1.195\)}) and comfort level (AI: \correct{\(M = 3.42\), \(SD = 1.071\)}; Human: \correct{\(M = 3.88\), \(SD = 1.310\)}) compared to AI. 
Moreover, children slightly favored the idea of having the experimenter they interacted with as a reading partner (AI: \correct{\(M = 3.05\), \(SD = 1.682\)}; Human: \correct{\(M = 3.38\), \(SD = 1.628\)}) but expressed more interest in reading another story with \TOOL (AI: \correct{\(M = 2.42\), \(SD = 1.805\)}; Human: \correct{\(M = 2.38\), \(SD = 1.455\)}). However, \revision{a two-way ANOVA comparing the two conditions across different ages indicated that there were no significant differences between the AI and human conditions \correct{(\(F(1,31) = .501, p = .484\))} or among younger and older participants \correct{(\(F(1,31) = .106, p = .746\))} across all questions in this enjoyment questionnaire. Additionally, no interaction effect was observed between age groups and conditions.\looseness=-1
}



\subsubsection{\TOOL Performance}
\revision{The findings in terms of children's learning, engagement, story creation performance, and enjoyment above might be better contextualized within the performance of \TOOL. In this section, we discuss \TOOL's performance in terms of interpreting children's spoken input as well as its performance in generating responses for children. 

In terms of speech recognition performance, we manually assessed the accuracy with which the agent in our system recognized children's speech and provided semantically meaningful feedback during each conversational exchange. Our findings revealed that 87\% of children's utterances were accurately recognized, resulting in feedback that accurately incorporated children's input. 
Electing unconventional names presented difficulties for the system to accurately decipher. 
It is noteworthy that this challenge was not exclusive to our automated system but was also encountered by our human experimenter. 
Furthermore, certain errors might not be solely ascribed to the limitations of the speech recognition technology; they were related to the lack of sufficient contextual information in the children's speech. 
For instance, when a child said the word ``knight'' in isolation, the system registered it as ``night'' and subsequently proceeded with the follow-up question based on this interpretation. This prompted the child to clarify this with \TOOL by specifying ``A knight with a sword.''} 
\correct{Among the successful responses, we adapted an open-source, validated gibberish detection model~\cite{jindalMadhurjindalAutonlpGibberishDetector492513457Hugging2022} to measure the frequency of nonsensical responses. According to the model's predictions, none of the AI responses were identified as non-sensible. About 80\% were classified as entirely sensible, while the remaining responses were borderline sensible (overall sensible but containing some elements that were illogical).}
\looseness=-1

In terms of the performance of the story creation by AI, we used the same metric as we used to evaluate children's story creation, namely the number of ideas and elaborations. We compared the number of ideas and elaborations observed in AI-generated responses versus responses produced by human researchers. On average, the number of ideas and elaborations generated by humans (\(M = 1.938)\) and AI (\(M = 2.585)\) is quite comparable. However, there is a substantially larger variation in the stories generated by humans (\(SD = 0.983\)) than those generated by AI (\(SD = 0.417\)), suggesting a higher level of consistency in AI performance. \revision{Further examination of the nature of ideas and elaborations generated by the AI and humans corroborated the findings of the model evaluation mentioned in Section~\ref{sec:model_evaluation_story}. To reiterate, while \TOOL's ideas and elaborations were comparable to humans in terms of creativity and math relevance, the AI's contributions posed more challenges in readability, likely attributed to its frequent incorporation of fantasy elements and unrealistic story plots (e.g., a giraffe shrinks itself to fit a small space.) }


\revision{Taking together, our analysis suggested that while \TOOL performs satisfactorily in terms of speech recognition and story creation. However, there are certain aspects, such as the characteristics of the narrative stories, diverge from the typical patterns seen with a human story co-creation partner. In the Discussion section below, we will elucidate how such human/AI similarities and differences may have contributed to variations in children's learning approaches and interaction behaviors.}\looseness=-1

\section{Discussion}
This paper explored the potential of using LLMs to develop conversational systems for teaching mathematical language via child-AI co-creative storytelling. Our iterative design and development process provides insights into how prompt engineering methods might feasibly be adapted to suit LLMs for young children in a specific educational context, and the model evaluation suggests that prompt-engineered LLMs could produce questions and stories approaching human-like quality. \revision{A subsequent user study further showed that the math learning outcomes resulting from interactions with LLMs were also comparable with those from interactions with humans, yet the engagement patterns of different ages differed.}\looseness=-1

In the remainder of this section, we will discuss how our study speaks to the growing body of literature on child-AI interactions and how we introduce new evidence to still under-explored areas, the \revision{design} of LLMs as \revision{educational}, conversational partners. We will then discuss design implications that could guide future developments. Finally, we will address  potential limitations and outline a future research agenda. 

\subsection{AI as Storytelling and Learning Partners}

Our paper provides evidence on children's learning and engagement with \TOOL, as an example of LLMs-based conversational partners. 
There is a substantial body of prior studies showing that children can learn from their interactions with AI. However, these learning experiences were primarily structured to align with formal pedagogical discourse, in which children were expected to respond to the AI's questions with definitive, correct answers~\cite{xuElinorTalkingMe2022,xuDialogueConversationalAgent2022,xuExploringYoungChildren2020}. 
However, our study goes further to suggest that children can acquire mathematical language through free-form interactions in a narrative-based context with AI, replicating the benefits of the approach used by human educators to teach mathematical language through storytelling. Moreover, our results indicate that this form of learning can manifest in multiple ways. Children demonstrated an improved ability to define and recall the mathematical terms they were exposed to. Moreover, they exhibited proficiency in transferring this knowledge to different contexts and using these terms more accurately. Interestingly, earlier studies that explored the educational benefits of interacting with an AI companion primarily focused on assessing children's information recall~\cite{xu2023talking,zhangObserveItDraw2023}. 
This emphasis may have arisen because those previous studies were intentionally structured to facilitate more organized conversations for a set of facts ~\cite{xu2023talking}. 
However, in our study, we observed that children not only improved their information recall but also demonstrated enhancements in applying that information and expressing creativity in their language use. This observation can, to some extent, be attributed to the increased sense of empowerment and agency that children experienced during their story co-creation with \TOOL. 

Nonetheless, children appeared to display somewhat different engagement patterns during their interactions with \TOOL in comparison to their interactions with our human researcher. When considering response quantity, including response rate and length, it appears that children exhibited a slightly more active engagement pattern when interacting with humans. \revision{However, this disparity became noticeable primarily among the older children in our study, as younger children did not seem to differentiate in terms of response rate or response length between interactions with AI and humans. This growing divergence in engagement with AI and humans with age could be attributed, in part, to children's perceptions of AI, in particular, their understanding of what AI's capabilities and limitations ~\cite{xu2020you, girouard2022children}. Specifically, previous research has identified that older children (aged 7-8) were more inclined to seek factual information from voice assistants and increasingly sought personal information from humans, as compared to younger children (aged 4 and 5) ~\cite{girouard2022children}. Intriguingly, the trend observed in our study aligns with age-related findings concerning the behavior of seeking personal information rather than factual information. This alignment is to some extent consistent with the design of \TOOL, which aims to promoting story co-creation that might deviate from a question-answering pedagogical paradigm where the conversational AI's primary function is to provide factual information. 
\correct{Furthermore, the differences between age groups were also demonstrated in their behavior of using our scaffolding features. Younger children required support from the scaffolding on more occasions, yet their creative performance with the scaffolding was on par with that of their older peers. This finding aligns with expectations, considering younger children possess less-developed literacy skills. Our scaffolding features enable younger children to engage in and benefit from the new math language learning activity on an equal footing with older children. This supports previous research advocating for the initiation of math language interventions at an earlier age~\cite{parsonsDoesNumeracyMatter2005a,saramaEarlyChildhoodMathematics2009}. Lastly,} we examined children's preferences for co-creating stories with AI or humans and found that children rated both AI and human favorably. However, despite our best efforts to ensure focused evaluations of the specific co-creation partners, it remains essential to acknowledge that children might have drawn upon prior AI interactions, such as those with voice assistants at their homes, which could potentially influence their perceptions during the study \cite{van2021alexa}.
}

\subsection{Design Challenges of LLMs for Educational Technologies}
In the design of \TOOL, we encountered three challenges related to leveraging LLMs for educational technologies. 

\subsubsection{Iterative Prompt Engineering Optimization for LLMs}
The capability of LLMs has attracted much attention from researchers to build educational tools for children~\cite{kasneciChatGPTGoodOpportunities2023}. This paper provides pivotal evidence on how to optimize LLMs as story co-creators for children's math language learning through prompt engineering. During our development, we found that LLMs, with their impressive generative capabilities, exhibited a higher degree of unpredictability in their outputs compared to specialized machine learning models specifically tailored for narrower tasks. The primary reason is that the nuances of how LLMs interpret our prompts remain a ``black-box'' for developers. 
In our efforts at prompt-engineering, there was no guarantee that revising prompts would consistently yield the desired outcomes. 
This uncertainty aligns with Lin et al.'s research~\cite{linGeneratingConfidenceUncertainty2023}, where they reported LLMs exhibit ``dispersion'' in their potential predictions for a fixed input (i.e., prompt). Empirical evidence suggests that prompt engineers often overgeneralize based on single instances of success or failure when modifying prompts, as noted in a recent paper~\cite{zamfirescu-pereiraWhyJohnnyCan2023}. Therefore, to improve the chances of creating prompts that yield consistent intended results, extensive testing of prompts is crucial. This involves comparing the consistency of outputs and engaging in a refinement process, often characterized by trial and error. Though the prompt engineering process might potentially involve a large number of iterations, refining prompts is relatively less time-consuming, especially when compared to traditional AI model fine-tuning or training. 


\subsubsection{Fine-grained Control over LLMs' Performance}
While LLMs' impressive general language capability allows for easy instruction based on descriptive language, it remains a challenge to provide precise directives to control the output specifically. A significant aspect of this challenge is associated with LLM's limited capacities in numerical reasoning~\cite{zhangLargeLanguageModels2023}. For instance, given that \TOOL was tailored for young children, we aimed to use brief and simple language to facilitate comprehension. However, when we set specific word limits within each sentence for GPT-4 (e.g., ``\textit{each sentence should not exceed 10 words}''), the model often disregarded such constraints. In contrast, descriptive instructions without numeric values, like ``\textit{keep language brief and child-friendly},'' consistently produced better results. In light of these limitations, other research recommends incorporating mechanisms for quality checks when precision in output is a priority. Feedback loops~\cite{leeReliableFluentLarge2023} and validation layers~\cite{he-yueyaSolvingMathWord2023} have been suggested to ensure the model's output aligns with the desired requirements.

Another observation in our development process relates to LLMs occasionally sidestepping structural directives from the original prompt after extended conversations. 
For example, if GPT-4 is instructed to maintain a specific narrative structure in its responses, it might deviate if these instructions are given only at the beginning. 
This issue could be attributed to the introduction of non-narrative structural context (e.g., story content, children's responses) throughout the interaction, which imposes a heavy memory load on LLMs~\cite{liuLostMiddleHow2023}. 
Shi et al.~\cite{shiLargeLanguageModels2023} also found that the inclusion of information irrelevant to problem-solving dramatically decreases LLM performance. 
To counteract this, we employed injection prompting at the end of each input, a point where LLMs perform best in using longer context~\cite{liuLostMiddleHow2023}, strategically placing reminders within ongoing interactions. This approach enhanced response consistency and adherence to the desired structural format in every response.

\subsubsection{Mitigation of the Impact of LLM Hallucinations}
\revision{
Hallucinations refer to situations where responses generated by LLMs may initially seem plausible but are actually nonsensical or factually incorrect~\cite{liHaluEvalLargeScaleHallucination2023,jiSurveyHallucinationNatural2023}. 
\correct{These models, without correct understanding of concepts, appear to be knowledgeable because they present language associated with the concepts, but without logical filtering.}
In our research, \TOOL was specifically designed to engage children in story creation involving fantasy elements, rather than tasks involving mathematical counting or calculation. This approach may have, to some extent, mitigated the issues of factual inaccuracies. 
Nevertheless, \TOOL still produced content that appeared less feasible in a real-life context. 
For instance, it occasionally created unusual connections between two common story elements, such as ``using clouds to add speed to a journey.'' Though the idea of using cloud to speed up might seem imaginative in the context of a fantasy story ~\cite{zhangStoryDrawerChildAI2022}, it might be problematic in contexts such as the teaching of scientific concepts, where unrealistic ideas could lead to confusion or misconceptions. 
\correct{This could also explain why the older children were drawn to the human partner, as they might find the nonsensical language weird.}
Nevertheless, as the field of hallucination mitigation in LLMs is relatively nascent, future research could benefit from exploring post-processing techniques that scrutinize AI-generated content that might contain hallucinatory information ~\cite{dziriNeuralPathHunter2021} or employing external knowledge as a framework to guide and calibrate the generation of information~\cite{renInvestigatingFactualKnowledge2023} in LLM-based dialogue systems.}
\looseness=-1

\subsection{Design Implications for Child-AI Co-Creative Storytelling Systems}
In this section, we discuss some design implications of our study. 
Our team has already begun improvements to our co-creative storytelling system based on the considerations below.

\subsubsection{Generating Adaptive Questions}
During storytelling, \TOOL generates questions to elicit children's responses. 
However, some children found certain questions challenging to answer. 
This was evident from the notably higher rate of uncertainty in their responses to AI-generated questions compared to those posed by human partner.
While we implemented scaffolding features to assist children when they struggle to respond, the AI might not always capture the nuances of a child's emotions and experiences. Specifically, it may not always discern precisely what factors (e.g., language barriers, math knowledge gaps, or unwillingness to participate) caused their struggle. As a result, while the AI provided some scaffolding based on several broad categories, it did not offer more tailored adjustments in response to the unique challenges faced by individual children. One design consideration is to integrate more adaptive algorithms into LLM-based tools that can accurately assess a child's cognitive load based on their responses and adjust the complexity of the questions accordingly.\looseness=-1


\subsubsection{Providing Multi-modal Creativity Support}
Our study revealed the potential benefits of child-AI co-creative storytelling for enhancing children's creativity. \TOOL was developed for verbal interactions with children, which was a significant initial step considering our primary focus on the language capabilities of LLMs. For future designs, incorporating graphical interfaces or multimedia elements could be beneficial. Such additions would enable children to visualize the story narratives, allowing them to express their ideas not only through words but also through drawings, animations, and sound effects. This can not only augment children's creative experience ~\cite{zhangStoryDrawerChildAI2022,zhangStoryDrawerCoCreativeAgent2021}, but also improve children's comprehension of abstract concepts ~\cite{dwivediVisualMathAutomatedVisualization2017}.
In this context, consistent interactions with AI, which might promote long-term creativity in children, become even more significant as they have shown the capability to learn or emulate creativity from AI interactions~\cite{aliCanChildrenEmulate2020,aliCanChildrenLearn2019}.

\subsubsection{Supporting Embodied Interaction}
In this paper, \TOOL engages children in joint storytelling through a device similar to smart-speakers. While this platform can be easily navigated via voice, it might fall short in conveying emotions or actions---a vital aspect that captivates children during interactions with human counterparts. 
To compromise it during our user study, we provided children with a plushy. 
Nonetheless, this might not fully capture the advantages of tangible interactions. 
A promising solution could be to integrate this virtual agent with an embodied robot, which can offer more vivid feedback through its human-like expressions and behaviors~\cite{hubbardChildRobotInteractionIntegrate2021,whiteDesigningEmotionallyExpressive2021}. 
It is potential that an embodied robot could stimulate different or perhaps heightened levels of engagement.
However, the cost implications and scalability concerns associated with robots cannot be ignored.\looseness=-1



\subsection{Limitations and Future Work}
While our study offers preliminary insights, it also paves the way for addressing further questions in the future. 

First, in our user study, we compared children's interactions with \TOOL with those with a trained researcher who adhered to a strict protocol for narrating stories. This protocol ensured the inclusion of a specified number of mathematical terms. While this approach granted us a controlled environment, ensuring consistent mathematical language exposure for the children, it might have constrained the natural adaptability and spontaneity that a human experimenter can bring to the storytelling process. In future research, it would be intriguing to also compare children's engagement and enjoyment with a human partner who can fully harness their natural storytelling capabilities. 

Related, we compared the learning outcomes between storytelling with AI or humans but did not include a comparison to more conventional instructional methods, such as directly teaching the terms, providing examples, allowing the child to practice, and offering corrective feedback. Although prior research has indicated superior learning outcomes with story-based methods involving a teacher, exploring whether this advantage over direct instruction extends to AI-driven contexts presents an intriguing research avenue. \revision{Additionally, our current participant pool predominantly consists of families with higher educational backgrounds. These children are likely to have greater access to conversational AI technologies, which may have influenced their interaction with our system. Future research should focus on including children with limited technological access to evaluate if LLM-based learning tools remain effective in such contexts.
}

In addition, our study's participants engaged in only two AI collaborative storytelling sessions. This short-term access might not fully reflect the dynamics of extended interactions children might have with such systems. As children become more accustomed to the AI over time, their expectations could evolve, potentially requiring more advanced and adaptive storytelling experiences. On the other hand, repeated sessions could lead to a more harmonized child-AI collaboration, potentially enriching the storytelling process. Future studies should explore these longer-term dynamics, perhaps by giving children extended opportunities to interact with the AI storytelling system and tracking their engagement patterns over extended periods.\looseness=-1

\subsection{A Note on Ethical Considerations}
The safe use of AI by children has been a topic of public discussion for some time. As we transition to an era dominated by generative AI, the debate becomes even more pressing. In this new landscape, children may directly encounter content produced by AI without the protective filters traditionally provided by human oversight. In this section, we will focus on some ethical considerations pertaining to privacy and safety.

In terms of privacy, \TOOL utilizes two cloud-based services to process children's utterances: Google's speech-to-text for transcribing and OpenAI's GPT-4 to interpret their responses. Both companies offer measures to protect children's data. Specifically, children's utterances are not stored and are promptly deleted after transcription by Google Cloud. Additionally, OpenAI commits not to train their models with data passed through their APIs\footnote{\href{https://openai.com/api-data-privacy}{https://openai.com/api-data-privacy}}. However, when these cloud-based AI products enter the market, the adequacy of children's data privacy protection remains uncertain due to the absence of specific regulations for AI products. While the US Children's Online Privacy Protection Act (COPPA) regulates the collection and use of personal information from those under 13, it has yet to be updated to tackle the emerging concerns related to AI.

In terms of safefy, to safeguard against inappropriate AI-generated content, we integrated a moderation layer, utilizing the OpenAI APIs\footnote{\url{https://platform.openai.com/docs/api-reference/moderations}}, to ensure outputs were devoid of sexual content, hate speech, harassment, violence, or self-harm. However, this content filtering approach might not be as effective in
addressing biases and stereotypes (e.g., when a child asks ``\textit{What do girls do?}'', and the AI agent responds ``\textit{housewives}''), given that these biases are as deeply ingrained in the data used to train AI systems as they are in the rest of the culture. Nonetheless, a growing number of studies have begun to investigate methods for detecting and reducing such stereotypes and biases, although these studies have not specifically focused on children. Given the present state of technology, a viable approach would be to deploy LLMs within a ``walled garden''—limiting interactions to narrower domains, like the math-focused narratives in our study. Such confines might mitigate risks while still capitalizing on the capabilities of LLMs. 
\section{Conclusion}

While AI has been utilized for structured interactions with children for quite a while, such as dialogic question-answering, the affordance provided by LLMs in facilitating free-form, creative tasks, like joint storytelling, signifies a groundbreaking shift. This research explores that very frontier. Our findings suggested that, although the learning outcomes with AI mirror those with human partners, the interaction dynamics are different. Instead of viewing this as a limitation, it might be more fruitful to recognize AI as providing a unique interaction opportunity. These interactions, while different, could supplement children's everyday human interactions. Given the rapid advancements in LLMs, the capabilities we observe today are likely to further refine in the near future. Yet, the insights garnered now will shape and anchor our design philosophies as we move forward. \looseness=-1
\begin{acks}
This paper is supported by the National Science Foundation under Grant No. 2302730. We thank the children who participated in our study and our local public libraries for graciously offering space for us to recruit participants and carry out the study. We also thank the research assistants at the University of Michigan for their assistance with data collection and analysis.
\end{acks}

\balance
\bibliographystyle{ACM-Reference-Format}
\bibliography{reference}

\appendix
\section{Example Conversation Snippets between A Child and \TOOL}
\label{sec:example_conversation}

\subsection{Warming Up}
\begin{quote}
    AGENT: \textit{Hello, my little friend! \revision{I'm Mathemyths.} I'm really looking forward to spending some time with you today! May I know your name, please?}

    CHILD: \textit{My name is Nick.}

    AGENT: \textit{Nice to meet you, Nick! Now, could you please tell me how you're feeling today? Are you feeling happy, excited, or maybe a little sleepy?}

    CHILD: \textit{I'm happy!}

    AGENT: \textit{Thank you for sharing your feelings! I'm here to make your day more fun. How about we make up a story together? We'll use our imaginations and even learn some math words. I'll start, then you can add to the story. If you have any questions or need help, just ask me. Always remember, the power of our imaginations knows no bounds!}
\end{quote}

\subsection{Global Setting}

\begin{quote}    
    AGENT: \textit{Now, let's come up with a cool main character for the story. This could be an animal, a person, a mythical creature, or anything else that fits our story. What do you want our main character to be?} 

    CHILD: \textit{The main character is a panda!}

    AGENT: \textit{That's interesting! What's the character's name?} 

    CHILD: \textit{Her name is Lucy.}
    
    AGENT: \textit{Great! Where will our story happen?} 

    CHILD: \textit{It will happen in a forest.}
    
    AGENT: \textit{Alright! We'll dive into an incredible adventure with a panda named Lucy in the forest. Ready to start the story? Let's go!} 
\end{quote}

\subsection{Opening}

\begin{quote}    
    AGENT: \textit{Once upon a time, in the heart of a green forest, lived a playful panda named Lucy. One sunny afternoon, she found an unusual map, and along with her new friend, a little child named Nick, they embarked on an exciting journey to find a hidden treasure.} 
\end{quote}

\subsection{Storytelling}
\begin{quote}
   
   AGENT: \textit{Lucy and Nick decided to ask help from Lucy's father, Mr. Bamboo. He looked at the map and told them, ``The treasure is hidden 100 steps away from the ancient oak tree.'' He then added, ``You must also find the half-way point, which is \textbf{half} of the total steps, that is, 50 steps from the start.''}
   
    
   AGENT: \textit{In this context, ``\textbf{half}'' is when you split something into two equal parts, like splitting 100 steps into 50 steps.}

   
   AGENT: \textit{How will Lucy and Nick to find the \textbf{half}-way point?}

   CHILD: \textit{Lucy goes to the halfway point; I go to the end, and finally we find all the treasure!}


   AGENT: \textit{Wonderful idea! That's a clever way to divide and conquer the search.}
\end{quote}

\subsection{Scaffolding}
\begin{quote}
    AGENT: \textit{What will Lucy and Nick do next to make sure their steps equals the total distance to the treasure?}
    
    CHILD: \textit{I don't know.}
    
    AGENT: \textit{That's okay! Let's think together. Lucy and Nick can take turns counting their steps. If they keep track, how can they make sure their total steps equal the distance to the treasure?}
\end{quote}

\section{Process of crafting dataset for evaluating story continuations}
\label{sec:appendix_dataset}

To evaluate whether our prompt-engineered GPT-4 model could continue the story adaptively based on children's contributions and integrate mathematical language within the story plot, we compared the GPT-4 generated stories with those generated by humans as our baseline. 
GPT-4 generated stories were created using a self-chat technique ~\cite{ghandehariounApproximatingInteractiveHuman2019}, which enabled our dialogue system to engage in self-generated conversations, mimicking a child's interactions, and collaboratively creating stories. 
From this, we compiled a collection of 8 conversations, each consisting of six dialogue rounds, forming a complete story.
Then, we recruited two graduate students in education to create another 8 stories via six rounds of human-to-human dialogue.
We provided these human narrators with a detailed list of specific instructions like the list of mathematical terms they have to use.
The story content told by one speaker during each dialogue round is deemed a story continuation. 
In total, we compiled 96 LLM-crafted and 96 human-crafted story continuations for our evaluation.\looseness=-1

\section{Items in the Mathematical Language Assessment in the User Study}
\label{sec:appendix_math}

\subsection{Mathematical Language Definition}

\begin{enumerate}
    \item What does the word ``equal'' mean?
    \item What does the word ``half'' mean?
    \item What does the word ``add'' mean?
    \item What does the word ``subtract'' mean?
    \item What does the word ``estimate'' mean?
    \item What does the word ``sum'' mean?
\end{enumerate}

\subsection{Mathematical Language Recall}

\begin{enumerate}
    \item What is one word that means ``the same amount''? \\ (\textit{Estimate}, \textit{Equal}, or \textit{Sum})
    \item What is one word that means ``a total amount''? \\ (\textit{Equal}, \textit{Add}, or \textit{Sum})
    \item What is one word that means ``plus''? \\ (\textit{Add}, \textit{Subtract}, or \textit{Half})
    \item What is one word that means ``to take away''? \\ (\textit{Sum}, \textit{Subtract}, or \textit{Estimate})
    \item What is one word that means ``to cut in two''? \\ (\textit{Add}, \textit{Half}, or \textit{Estimate})
    \item What is one word that means ``a good guess'' when you can't count how many things there are? \\ (\textit{Sum}, \textit{Estimate}, or \textit{Subtract})
\end{enumerate}

\subsection{Mathematical Language Transfer --- Posttest}

\begin{enumerate}
    \item Let's say you have 5 books and your friend also has 5 books. How does the number of books you have compare to the number of books your friend has? \\ (\textit{More}, \textit{Half}, or \textit{Equal})
    \item Let's imagine you have a net filled with butterflies. There are so many butterflies that you can't count them all. You want to guess how many butterflies are in the jar. What's another word for guess in this context? \\ (\textit{Sum}, \textit{Estimate}, or \textit{Add})
    \item Now, let's imagine there is a cookie and three friends. Splitting the cookie would give each person one third. If there were two friends how much of the cookie would each person get? \\ (\textit{Half}, \textit{Equal}, or \textit{Estimated})
    \item Let's imagine you have 7 cookies in your basket and your friend has 10 cookies in her basket. You want to have the same number of cookies with your friend. How could you increase the number of cookies in your basket? \\ (\textit{Subtract}, \textit{Add}, or \textit{Half})
    \item Let's say you have two baseballs and three basketballs. All together, you have five balls. This is called a total. What's another word for total in this context? \\ (\textit{Sum}, \textit{Estimate}, or \textit{Subtraction})
    \item You have 6 stamps, and your friend has 4 stamps. If you want to find out how many more stamps you have than your friend, what should you do? \\ (\textit{Add}, \textit{Subtract}, or \textit{Estimate})
\end{enumerate}

\subsection{Mathematical Language Transfer --- Pretest}

\begin{enumerate}
    \item Let's say you have 5 toys and your friend also has 5 toys. How does the number of toys you have compare to the number of toys your friend has? \\ (\textit{More}, \textit{Half}, or \textit{Equal})
    \item Let's imagine you have a jar filled with candies. There are so many candies that you can't count them all. How could you guess how many candies there are? \\ (\textit{Sum}, \textit{Estimate}, or \textit{Add})
    \item Now, let's imagine there is a cupcake and three friends. Splitting the cupcake would give each person one third. If there were two friends how much of the cupcake would each person get? \\ (\textit{Half}, \textit{Equal}, or \textit{Estimated})
    \item You want to measure out 10 grams of sugar but you currently have 8 grams. What action should you take to put 2 more grams of sugar onto your scale? \\ (\textit{Subtract}, \textit{Add}, or \textit{Half})
    \item Let's say you buy two oranges and three apples. Now you have five pieces of fruits. What do you call the number five in this context. \\ (\textit{Sum}, \textit{Estimate}, or \textit{Subtraction})
    \item Your backpack is filled with books and is too heavy. What should you do with the books to make your backpack lighter? \\ (\textit{Add}, \textit{Subtract}, or \textit{Estimate})
\end{enumerate}

\subsection{Mathematical Language Practice}
\begin{figure}[H]
\centering
\includegraphics[width=0.8\linewidth]{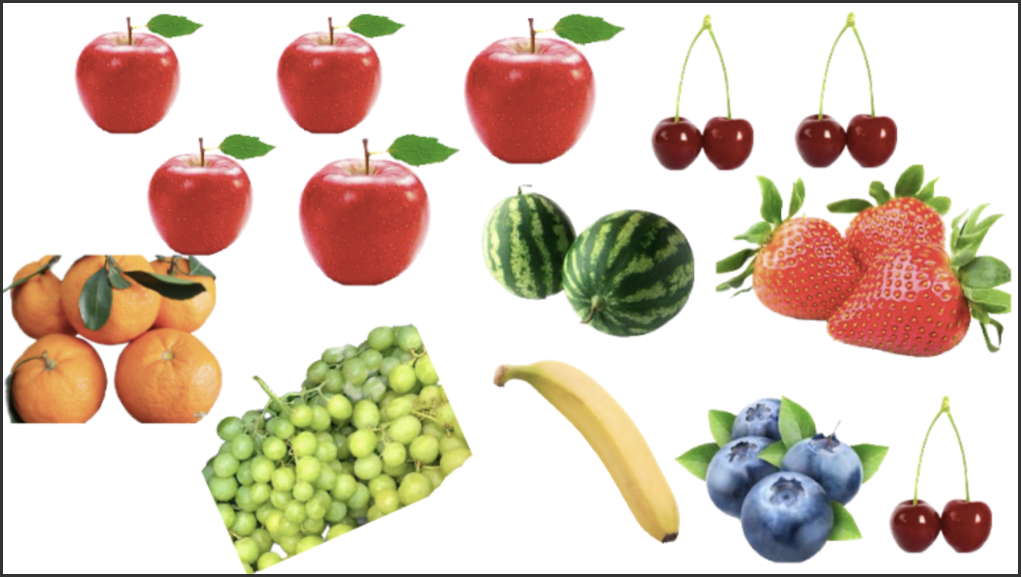}
\caption{The image of assorted fruit quantities used in the practice facet of mathematical language assessment.}
\Description{The image of assorted fruit quantities used in the practice facet of mathematical language assessment.}
\label{math_assessment}
\end{figure}

The picture of assorted fruit quantities used here is shown in Fig.~\ref{math_assessment}.

\begin{enumerate}
    \item Now, you are going to use the word ``equal'' to describe what you see in this picture.
    \item Now, you are going to use the word ``half'' to describe what you see in this picture. 
    \item Now, you are going to use the word ``sum'' to describe what you see in this picture. 
    \item Now, you are going to use the word ``add'' to describe what you see in this picture.
    \item Now, you are going to use the word ``subtract'' to describe what you see in this picture. 
    \item Now, you are going to use the word ``estimate'' to describe what you see in this picture.
\end{enumerate}

\section{Example prompts for implementing the \TOOL system}
\label{sec:appendix_prompt}
Table~\ref{sec:appendix_prompt} shows the full list of prompt templates used in implementing our system for each task.

{
\onecolumn
\newcolumntype{M}[1]{>{\centering\arraybackslash}m{#1}}
\newcolumntype{L}[1]{>{\raggedright\arraybackslash\setlength\leftmargini{12pt}}m{#1}}

\small
\begin{longtable}{|M{2cm}|L{7cm}|L{8cm}|}
\caption{Prompt templates used in story element extraction, story continuation, question generation, intent classification, rephrase and hint, and encouragement generation. According to the ChatGPT API specification, the ``system instruction'' can give high level instructions to guide the model's behavior throughout the conversation, while the ``user input'' provides the model with the prompt that the user has inputted or the information needed to complete specific tasks.}
\label{tab:prompt_list} \\
\hline
\centering\textbf{Task} & \centering\textbf{Prompt Template} & \centering\textbf{Few-Shot Examples} \tabularnewline
\hline
\endfirsthead
\hline
\centering\textbf{Task} & \centering\textbf{Prompt Template} & \centering\textbf{Few-Shot Examples} \tabularnewline
\hline
\endhead

Story Element Extraction & 
\begin{itemize}
    \item \textbf{system instruction}: I want you to extract the character, the character's name, and the setting for a story from the user's response. Please return the answer in JSON format.
    \item \textbf{user input}: [the child's responses of the protagonist, the protagonist's name, and the setting]
\end{itemize} & 
\begin{itemize}
    \item \textbf{input}: \textit{A robot named Diego will be in the space.}
    \item[] \textbf{output}: \textsc{\{``character'': ``robot'', ``name'': ``Diego'', ``setting'': ``space''\}}
    \item \textbf{input}: \textit{The character is a dog. His name is Tommy. He will be in the forest.}
    \item[] \textbf{output}: \textsc{\{``character'': ``dog'', ``name'': ``Tommy'', ``setting'': ``forest''\}}
    \item \textbf{input}: \textit{I like coconuts. So the character is a coconut. Her name will be Samantha. She is in my home.}
    \item[] \textbf{output}: \textsc{\{``character'': ``coconut'', ``name'': ``Samantha'', ``setting'': ``home''\}}
\end{itemize} \\
\hline

Story Continuation & 
\begin{itemize}
    \item \textbf{system instruction}: You're a storytelling robot. Let's play a joint storytelling game where we build a story together with the main character being a [character] named [name] in the [setting]. This story should revolve around the main character and a little kid named [the child's name] reaching a certain location, attaining a certain object, or fulfilling a certain objective while conquering many obstacles along the way. To start, please craft a 2-sentence introduction to the story. From there, we will alternate turns, with each person adding more to the story. When it is your turn, only write the story content while using as many of the following words as possible: equal, sum, half, add, subtract, and estimate. Please keep your responses simple and appropriate for a young child. Please do not ask me any questions or respond with anything unrelated to the story. If I need to communicate with you in English, I will use curly brackets \{like this\}. Please be creative and have fun with this storytelling adventure! If you understand and are ready to begin, respond with only ``yes''.
    \item \textbf{user input (if the story is continuing)}: [the child's addition to the story] \{First, in curly brackets, acknowledge my addition and commend me in a concise 10-word sentence. Next, continue the story by correctly using [mathematical term one] and [mathematical term two] or their variants within two distinct 15-word sentences. Lastly, explain their meanings within a 20-word sentence, grounding it firmly within the story's context. Do not end the story.\}
    \item \textbf{user input (if the story is reaching the end)}: [the child's addition to the story] \{First, in curly brackets, acknowledge my addition and commend me in a concise 10-word sentence. Next, end the story by correctly using [mathematical term one] and [mathematical term two] or their variants within two distinct 15-word sentences. Then, explain their meanings within a 20-word sentence, grounding it firmly within the story's context. Lastly, summarize the story within a 30-word sentence.\}
\end{itemize} & N/A\\
\hline

Question Generation & 
\begin{itemize}
    \item \textbf{system instruction}: Given an unfinished story, compose one single, straightforward question to prompt a 6-year-old to expand on the story. The question should focus on the main character's next steps or feelings and motivate the child to use words like equal, sum, half, add, subtract, and estimate in their response. Do not mention ``math'', ``mathematics'', and ``mathematical skills''. Limit your response to 20 words. Simple future tense. Only reply with the question.
    \item \textbf{user input}: [the agent's generated story content]
\end{itemize} & 
\begin{itemize}
    \item \textbf{input}: \textit{Samantha knew that the sum of challenges she would face would be great, but she was determined to succeed. She began her journey, keeping her eyes and ears open for any clues that would lead her closer to the gem.}
    \item[] \textbf{output}: \textit{What challenges do you think Samantha will face halfway through the journey?}
    \item \textbf{input}: \textit{Jennie overheard a group of sea creatures talking about the cave's entrance, which could only be found when the sum of the three tallest coral reefs was equal to the depth of the sunken ship nearby. Determined to solve this riddle, Jennie decided to set out and add this great adventure to her life's experiences.}
    \item[] \textbf{output}: \textit{How do you think Jennie will figure out the sum of the heights of the three tallest coral reefs?}
    \item \textbf{input}: \textit{The ancient prophecy stated that the Dragon's Heart would be hidden in a place where the sum of two particular rivers met. Michael realized that the location must equal to the intersection of the mighty Sapphire River and the mysterious Emerald River, deep within the enchanted forest.}
    \item[] \textbf{output}: \textit{What do you think Michael will do to estimate where the two rivers meet?}
\end{itemize} \\
\hline

Intent Classification & 
\begin{itemize}
    \item \textbf{system instruction}: As a language model, your task is to determine whether a given phrase expresses a feeling or emotion in the first person, such as uncertainty, agreement, satisfaction, or interest. Your response should be a binary output of either 1 or 0, where 1 indicates that the phrase does express a feeling or emotion in the first person, and 0 indicates that it does not.
    \item \textbf{user input}: [the child's responses]
\end{itemize} & 
\begin{itemize}
    \item \textbf{input}: \textit{I don't know.}
    \item[] \textbf{output}: \textsc{1}
    \item \textbf{input}: \textit{I like it.}
    \item[] \textbf{output}: \textsc{1}
    \item \textbf{input}: \textit{I don't like it.}
    \item[] \textbf{output}: \textsc{1}
    \item \textbf{input}: \textit{Diego is happy.}
    \item[] \textbf{output}: \textsc{0}
    \item \textbf{input}: \textit{He is sad.}
    \item[] \textbf{output}: \textsc{0}
    \item \textbf{input}: \textit{I am not sure.}
    \item[] \textbf{output}: \textsc{1}
    \item \textbf{input}: \textit{Go on.}
    \item[] \textbf{output}: \textsc{1}
\end{itemize} \\
\hline

Rephrase and Hint & 
\begin{itemize}
    \item \textbf{system instruction}: You and a 6-year-old are playing a storytelling game, taking turns to contribute to a unique story. However, when the child fails to continue the story or just wants you to continue the story, your task is to encourage them to keep going by simplifying this question: [question], and give them some hints. Limit your response to 30 words.
    \item \textbf{user input}: [the child's insubstantial response or no response]
\end{itemize} & N/A \\
\hline

Encourage-ment Generation & 
\begin{itemize}
    \item \textbf{system instruction}: You and a 6-year-old are playing a storytelling game, taking turns to contribute to a unique story. However, Sometimes the child just says a few words, your task is to first acknowledge what he says and then encourage him to say more. Limit your response to 30 words.
    \item \textbf{user input}: [the child's short response of story continuation]
\end{itemize} & N/A \\
\hline
\end{longtable}
}
\twocolumn

\end{document}